\documentstyle[epsfig,12pt]{article}
\newlength{\dinwidth}
\newlength{\dinmargin}
\newcommand{\epsl}{\varepsilon \hspace{-5pt} / }

\newcommand{\qsl}{q \hspace{-5pt} / }

\newcommand{\ba}{\begin{array}}
\newcommand{\ea}{\end{array}}
\newcommand{\be}{\begin{equation}}
\newcommand{\ee}{\end{equation}}
\newcommand{\bea}{\begin{eqnarray}}
\newcommand{\eea}{\end{eqnarray}}

\def\bra{\langle}
\def\ket{\rangle}

\def\a{\alpha}
\def\b{\beta}
\def\g{\gamma}

\def\e{\epsilon}
\def\eir{\epsilon_{ir}}
\def\p{\pi}

\def\ep{\varepsilon}

\def\l{\lambda}
\def\m{\mu}
\def\n{\nu}
\def\G{\Gamma}

\def\mw{m_W}
\def\mt{m_t}
\def\mb{m_b}
\def\mwt{\mu_{Wt}}
\def\mub{\mu_b}

\def\Li{\mbox{Li}(1-\frac{1}{z})}
\def\to{\rightarrow}
\begin{document}
\thispagestyle{empty}
\addtocounter{page}{-1}
\begin{flushright}
DESY 97-040 \\
ITB-SB-97-18\\
hep-ph/9703349\\
March 1997
\end{flushright}
\vspace*{1.8cm}
\centerline{\Large\bf Two-loop  matching}
\vspace*{0.2cm}
\centerline{\Large\bf  of the dipole operators for 
$b \to s \gamma$ and $b \to s g$ 
\footnote{Work supported in part by Schweizerischer
Nationalfonds}}
\vspace*{2.0cm}
\centerline{\large\bf Christoph Greub }
\vspace*{0.4cm}
\centerline{\large\it Deutsches Elektronen-Synchrotron DESY,}
\centerline{\large\it 22603 Hamburg, Germany}
\vspace*{0.8cm}
\centerline{\large\bf Tobias Hurth}
\vspace*{0.4cm}
\centerline{\large\it Institute of Theoretical Physics, SUNY at Stony Brook,}
\centerline{\large\it Stony Brook, New York 11794-3840, USA}
\vspace*{2.5cm}
\centerline{\Large\bf Abstract}

\vspace*{1.5cm}
The order $\a_s$ corrections 
to the Wilson coefficients  
of the dipole
operators ($O_7,O_8$) at the matching scale $\mu =m_W$ are a crucial
ingredient for a complete next-to-leading logarithmic
calculation of the branching ratio for $b \to s \gamma$.
Given the phenomenological relevance and the fact that
this two-loop calculation has been done so far only by
one group \cite{Adel},
we present a detailed re-calculation using a different method.
Our results are in complete agreement with those in ref. \cite{Adel}.

\newpage
\section{Introduction}
\label{sec:introd}
\setcounter{equation}{0}
By definition, rare $B$ meson decays only arise at the one loop
level in the standard model (SM). Therefore these decays are 
particulary sensitive to effects from new physics.
Among these decays, the inclusive modes like $B \to X_s \gamma$
are particulary interesting, because no specific model is needed
to describe the final hadronic state in contrast to the exclusive
decay modes. Indeed, heavy quark effective theory tells us that
the decay width $\G(B \to X_s \gamma)$ 
is well approximated by the partonic decay 
rate $\G(b\to X_s \gamma)$ which can be
analyzed in renormalization group improved perturbation theory. 
The class of 
non-perturbative effects which scales like $1/m_b^2$
is expected to be well below $10\%$ \cite{Falk}. This numerical
statement is supposed to hold also 
for the recently discovered non-perturbative contributions
which scale like $1/m_c^2$ \cite{Wise}. 

Up to recently, only the leading logarithmic (LL) perturbative QCD
corrections were calculated {\it systematically}
\cite{counterterm}. The error of these
calculations is dominated by a large 
renormalization scale dependence at the $\pm 25\%$ level. 
The measured branching ratio
$BR(B \to X_s \g)=(2.32 \pm 0.67) \times 10^{-4}$ reported in 1995
by the CLEO collaboration \cite{Cleo2} overlaps with the estimates
based on leading logarithmic calculations
(or with some next-to-leading effects partially included)
and the experimental and 
theoretical errors are comparable 
\cite{AG91,AG93,Burasno,Ciuchini,AG95,Shifman}. 
However, in view of the expected 
increase in the experimental precision in the near future, 
it became clear that 
a systematic inclusion of the next-to-leading logarithmic (NLL) corrections
becomes necessary \cite{Burasno}. 
This ambitious NLL enterprise was recently completed; 
combining the results  of different groups 
\cite{Adel,AG91,AG95,Pott,GHW,Mikolaj}, 
the first complete theoretical prediction to NLL  pecision 
for the $b \to X_s + \gamma$ branching ratio 
was presented in \cite{Mikolaj}:
$BR(B \to X_s \g)=(3.28 \pm 0.33) \times 10^{-4}$.
This prediction is still in agreement with the CLEO
measurement at the $2\sigma$-level. The theoretical error is twice smaller
than in the leading logarithmic prediction. 
So the inclusive $B \to X_s + \gamma$ 
mode will provide an interesting test of the SM and its extensions
when also more precise experimental data will be available.

Before discussing in some more detail the principle steps 
leading to a next-to-leading
result for $b \to X_s \gamma$, we briefly have to recall the formalism.
We use the framework of an effective 
low-energy theory with five quarks, obtained by integrating out the
 top quark and the $W$-boson. 
The effective Hamiltonian relevant for $b \to s \gamma$ and
$b \to s g$ reads 
\begin{equation}
\label{heff}
H_{eff}(b \to s \gamma)
       = - \frac{4 G_{F}}{\sqrt{2}} \, \lambda_{t} \, \sum_{i=1}^{8}
C_{i}(\mu) \, O_i(\mu) \quad ,
\end{equation}
where $O_i(\m)$ are the relevant operators,
$C_{i}(\mu)$ are the corresponding Wilson coefficients,
which contain the complete top- and W- mass dependence,
and $\lambda_t=V_{tb}V_{ts}^*$ with $V_{ij}$ being the
CKM matrix elements \footnote{The CKM dependence globally factorizes,
because we work in the approximation $\l_u=0$.}.
Neglecting operators with dimension $>6$ which are suppressed 
by higher powers of $1/m_{W/t}$-factors and using the equations
of motion for the operators, one arrives at the following basis 
\footnote{In \cite{Mikolaj} another basis was used. We comment on this
in the summary.} 
of dimension 6 operators \cite{Grinstein90}
\bea
\label{operators}
O_1 &=& \left( \bar{c}_{L \b} \g^\m b_{L \a} \right) \,
        \left( \bar{s}_{L \a} \g_\m c_{L \b} \right)\,, \nonumber \\
O_2 &=& \left( \bar{c}_{L \a} \g^\m b_{L \a} \right) \,
        \left( \bar{s}_{L \b} \g_\m c_{L \b} \right) \,,\nonumber \\
O_3 &=& \left( \bar{s}_{L \a} \g^\m b_{L \a} \right) \, \left[
        \left( \bar{u}_{L \b} \g_\m u_{L \b} \right) + ... +
        \left( \bar{b}_{L \b} \g_\m b_{L \b} \right) \right] \,,
        \nonumber \\
O_4 &=& \left( \bar{s}_{L \a} \g^\m b_{L \b} \right) \, \left[
        \left( \bar{u}_{L \b} \g_\m u_{L \a} \right) + ... +
        \left( \bar{b}_{L \b} \g_\m b_{L \a} \right) \right] \,,
        \nonumber \\
O_5 &=& \left( \bar{s}_{L \a} \g^\m b_{L \a} \right) \, \left[
        \left( \bar{u}_{R \b} \g_\m u_{R \b} \right) + ... +
        \left( \bar{b}_{R \b} \g_\m b_{R \b} \right) \right] \,,
        \nonumber \\
O_6 &=& \left( \bar{s}_{L \a} \g^\m b_{L \b} \right) \, \left[
        \left( \bar{u}_{R \b} \g_\m u_{R \a} \right) + ... +
        \left( \bar{b}_{R \b} \g_\m b_{R \a} \right) \right] \,,
        \nonumber \\
O_7 &=& (e/16\p^{2}) \, \bar{s}_{\a} \, \sigma^{\m \n}
      \, (m_{b}(\mu)  R + m_{s}(\mu)  L) \, b_{\a} \ F_{\m \n} \,,
        \nonumber \\
O_8 &=& (g_s/16\p^{2}) \, \bar{s}_{\a} \, \sigma^{\m \n}
      \, (m_{b}(\mu)  R + m_{s}(\mu)  L) \, (\l^A_{\a \b}/2) \,b_{\b}
      \ G^A_{\m \n} \quad .
        \nonumber \\
\eea

 In the dipole type operators $O_7$ and $O_8$, 
$e$ and $F_{\m \n}$ ($g_s$ and $G^A_{\m \n}$)
denote the electromagnetic (strong)
coupling constant and field strength tensor, respectively. 

%
It is well-known that the QCD corrections enhance the $b \to s \gamma$
decay rate by more than a factor of two; 
these QCD effects can be attributed to logarithms of the form 
$\alpha_s^n(m_b) \, \log^m(m_b/M)$,
where $M=m_t$ or $M=m_W$ and $m \le n$ (with $n=0,1,2,...$).
Working to NLL precision means, that one is resumming all the
terms of the form $\a_s^n(\mb) \, \ln^n (\mb/M)$, as well as
$\a_s(\mb) \, \left(\a_s^n(\mb) \, \ln^n (\mb/M)\right)$.
This is achieved by performing the following 3 steps:
%

\begin{description}
\item[Step 1] One has to match the full standard model theory
with the effective theory at the scale $\m=\mwt$, where
$\mwt$ denotes a scale of order $m_W$ or $m_t$. At this scale,
the matrix elements of the operators  in the 
effective theory lead to the  same logarithms  as the full theory
calculation. 
Consequently, the Wilson coefficients 
$C_i(\mwt)$ only pick up small QCD corrections,
which can be calculated in fixed order perturbation theory.
In the NLL program, the matching has to be worked out at the 
$O(\a_s)$ level. 
\item[Step 2] Then one performs the   
evolution of these Wilson coefficients from 
$\m=\mwt$ down to $\m = \mub$, where $\mub$ is of the order of $m_b$.
As the matrix elements of the operators evaluated at the low scale
$\mub$ are free of large logarithms, the latter are contained in resummed
form in the Wilson coefficients. For a NLL calculation, this RGE step
has to be performed using the anomalous dimension matrix up 
to order $\a_s^2$.
\item[Step 3]
The corrections to the matrix elements 
of the operators $\bra s \g |O_i (\mu)|b \ket$ at the scale  $\mu = \mub$
have to be calculated to order $\a_s$ precision.
\end{description}
The most difficult part in Step 1 is the 
 two-loop (or  
order $\a_s$) matching of the dipole operators, 
which has been worked out by Adel and Yao \cite{Adel} some time ago.
Step 3 basically consists of Bremsstrahlung corrections and virtual
corrections. The Bremsstrahlung corrections, 
together with some virtual corrections needed to cancel
infrared singularities, have been worked
out by Ali and Greub \cite{AG91,AG95}; later, this part
was confirmed and extended by \cite{Pott}. Recently, a  
complete analysis of the virtual corrections (up to the contributions 
of the 4 Fermi operators with very small coefficients) were presented
by Greub, Hurth and Wyler \cite{GHW}.     
The main result of the latter  analysis
consists in a drastic reduction of the renormalization 
scale uncertainty from about $\pm 25\%$ to about $\pm 6\%$.
Moreover, the central value was shifted outside the $1\sigma$ bound of the
CLEO measurement. However, at that time, the essential
coefficient $C_7(\mub)$ was only known to leading-log precision. 
It was therefore
unclear, how much the overall normalization will be changed, when
using the NLL value for $C_7(\mub)$. 
Very recently, the order $\a_s^2$ anomalous matrix (step 2) has been 
completely worked out
by Chetyrkin, Misiak and M\"unz \cite{Mikolaj}.
Using the matching result of Adel and Yao, these authors got 
the next-to-leading result for $C_7(\mub)$.  
 
Numerically, the LL and the NLL value 
for $C_{7}(\mub)$ 
are rather similar; the NLL 
corrections to the Wilson coefficient $C_7(\mub)$ 
lead to a change of the  $b \to X_s \gamma$
decay rate which does not exceed  6\% \cite{Mikolaj}: 
 The new contributions can be split into a part
which is due to the order $\a_s$ corrections to the matching (Step 1) 
and into
a part stemming from the improved anomalous dimension matrix (Step 2). 
While individually these two parts are not so small (in the
NDR scheme, which was used in \cite{Mikolaj}), they almost cancel
when combined as illustrated in \cite{Mikolaj}. 
This shows that all the different pieces are numerically
equally important. However, strictly speaking
the relative importance of different 
NLO-corrections at the scale $\mu = \mub$, namely 
the order $\alpha_s$ corrections to the matrix elements of 
the operators (Step 3) and the improved
Wilsoncoefficients $C_i$ (Step 1+2), 
is a renormalization-scheme dependent issue; 
so we stress that the discussion  
above was done within  
the naive dimensional regularization scheme (NDR).   

Each of the three steps implies  rather involved computations:
The calculation of the matrix elements (Step 3) involves two-loop diagrams
where the full charm mass
dependence has to be taken into account.
Also the matching calculation (Step 1) involves two-loop diagrams 
both in the 
full and in the effective theory.
Finally, the extraction of some of the elements in the $O(\a_s^2)$ anomalous 
dimension 
matrix involves three-loop diagrams.  
Given the fact, that it took a rather long time until 
the leading logarithmic calculations performed by different groups
converged to a common answer, it is certainly 
desirable that all three steps mentioned above 
should be repeated by other independent groups, 
and, may-be using other methods.

Making a step into this direction, we present 
in this paper a re-calculation of the two-loop 
matching of the dipole operators $O_7$ and $O_8$.
We extracted the $O(\a_s)$ contributions of the corresponding
Wilson coefficents $C_7$ and $C_8$ by calculating the on-shell
processes 
$b \to s \g$ and $b \to s g$
in both versions of the theory up to order $\a_s$. 
We worked out the
two-loop integrals by using   
the Heavy Mass Expansion method \cite{HMEsum}, 
which we describe in some detail in section 2.4. 
 
The rest of the paper is organized as follows. In section 2 
we make some preparations for the two-loop calculations.
We first 
explain how to extract
the order $\a_s$ corrections to the Wilson coefficients 
$C_7(\mwt)$ and $C_8(\mwt)$ in principle. Then,
in various subsections we discuss and illustrate
the technical methods used.
Sections 3, 4 and 5 are devoted to the computation of $C_{71}(\mwt)$:
In section 3 we calculate QCD corrections to $b\to s\g$ in the full
theory together with the corresponding counterterm contributions,
while in section 4 the same is done in the effective theory.
Comparing the results from section 3 and section 4, we extract 
$C_{71}(\mwt)$ in section 5. Similarly,
sections 6, 7 and 8 are devoted to the computation of $C_{81}(\mwt)$:
In section 6 we calculate QCD corrections to $b\to s g$ in the full
theory together with the corresponding counterterm contributions,
while in section 7 the same is done in the effective theory.
Comparing the results from section 6 and section 7, we extract 
$C_{81}(\mwt)$ in section 8. 
Finally, we give a brief summary in section 9.

\section{Preparations for the two-loop calculations}
\label{sec:principle}
\setcounter{equation}{0} 

\subsection{Strategy for extracting $C_{71}$ and $C_{81}$}
Let $\hat{M}$ denote the (on-shell) $b \to s \g$ 
matrix element calculated
in the effective theory. $\hat{M}$ can be written in the form
\be
\label{mhatform}
\hat{M} = \sum_i \, C_i (\m) \, \bra O_i(\m) \ket \quad , \quad
\bra O_i (\m)\ket \equiv \bra s \g|O_i(\m)|b \ket \quad .
\ee
To keep the notation simpler, we denote the matching scale
by $\m$ instead of $\mwt$.
Making use of the $\a_s$ expansion for $C_i(\m)$ and $O_i(\m)$
\be
\label{ciexp}
C_i(\m) = C_{i0}(\m) + \frac{\a_s}{4\pi} \, C_{i1}(\m) + \ldots \quad ,
\quad
\bra O_i(\m) \ket = \bra O_i(\m) \ket_0 +
\frac{\a_s}{4\p} \, \bra O_i(\m) \ket_1 + \ldots \quad , 
\ee
we get the corresponding expansion for $\hat{M}$ in the form
\be
\label{mhatexp}
\hat{M} = C_{i0}(\m) \, \bra O_i(\m) \ket_0 + \frac{\a_s}{4\p} \,
\left( C_{i0}(\mu) \, \bra O_i(\m) \ket_1 +
       C_{i1}(\mu) \, \bra O_i(\m) \ket_0 \, \right) + \ldots \quad . 
\ee

On the other hand, let $M$  denote the $b \to s \g$ matrix element evaluated
in the full theory after discarding power supressed terms of order 
$1/m_{W/t}^3$; $M$ has the expansion
\be
\label{expo}
M = M_0 + \frac{\a_s}{4\p} \, M_1 + \ldots \quad .
\ee
Requiring  $M=\hat{M}$ and taking the coefficient of $\a_s^1$,
we get the $O(\a_s)$ matching condition
\be
\label{matchcond}
M_1 =  C_{i0}(\mu) \, \bra O_i(\m) \ket_1 +
       C_{i1}(\mu) \, \bra O_i(\m) \ket_0  \quad . 
\ee 
All coefficients in eq. (\ref{matchcond}) are known \cite{Buras,Ciuchini2},
except \footnote{Of course $C_{71}$ and $C_{81}$ are also known from
Adel and Yao \cite{Adel}, but this is what we want to check.}
$C_{71}$ and $C_{81}$. 
As $C_{81}$ comes together with 
$\bra s\g |O_8(\m)|b\ket_0$, which is zero, eq. (\ref{matchcond})
has only one unknown, viz. $C_{71}$, i.e., just what we want to extract.

The discussion for the extraction of $C_{81}$ goes 
exactly along the same lines,
using the process $b \to s g$ instead of $b \to s \gamma$.    

A general remark is in order here. 
One could also match off-shell Greens functions instead of
on-shell matrix elements. However, in this case 
one is not allowed to work in the operator basis
given in eq. (\ref{operators}), because one has used the equations
of motion for the operators to get this 8 dimensional basis. This 
Hamiltonian
therefore only reproduces on-shell matrix elements correctly
\cite{Politzer}.
As we would have to work in the off-shell basis when matching Greens
functions,
we preferred to do on-shell matching.
There is of course a price to pay: The on-shell processes $b \to s \g$
and $b \to s g$ are plagued with infrared singularities, which have to be
treated carefully. However, as we will see later, this is not
a real problem.

\subsection{Technical details}
We work in 
$d=4-2\e$ dimensions; in the full theory 
we use anticommuting $\g_5$, which 
should not be a problem, because there are
no closed fermion loops involved. 
We also use this naive dimensional regularization scheme (NDR)
in the effective theory.
The calculations are done in the  `t Hooft-Feynman gauge 
(electroweak sector) and 
the gluon propagator is taken in the Feynman gauge. 
To avoid Euler $\g_E$ terms and $\ln(4\pi)$ factors 
in our expressions, we introduce the renormalization scale in the form 
$\mu^2 \, \exp(\g_E)/(4\p)$ ($\overline{MS}$ subtraction then corresponds
to subtracting the poles in $\e$).
Besides the ultraviolet singularities 
also the infrared singularitites are dimensionally regularized.
As we could clearly separate infrared and ultraviolet singularities,
we labeled the infrared poles by the index $ir$ (e.g., $1/\eir$).
We put $m_s=0$, except in situations where mass singularities 
appear, i.e., we
treat $m_s$ as a regulator of these singularities. 
We work in the approximation $\lambda_u=0$. To keep the formulae
more compact, we put immediately $Q_u=2/3$ ($Q_d=-1/3$) for up-type 
(down-type) quark charges.  
For the same reason we also immediately insert the numerical values 
for the color factors in the $b \to s g$ case. 

\subsection{Reducing the number of diagrams}
For reasons of gauge invariance, we know that the final result
for the $b \to s \gamma$ matrix element
can be written in the form
\be
M (b\to s \g) = F(\mbox{masses, couplings}) \quad 
 \bra s \g |O_7|b \ket_{tree}
\quad .
\ee
For $m_s=0$, the quantity $ \bra s \g |O_7|b \ket_{tree} $  is given by
\be
\label{decomp}
\frac{16 \, \pi^2}{e} \,
\bra s \g |O_7|b \ket_{tree} = 2  \mb \,\bar{u}(p') \, \epsl \, \qsl \, R \,
u(p) = \bar{u}(p') \, \left[ \,2 \, m_b^2 \, 
\epsl  \, L - 4 \, m_b \, (p\ep) \, R \, \right] \, u(p) \quad ,  
\ee
where $u(p')$ and $u(p)$ are the Dirac spinors
for the $s$ and the $b$ quarks, respectively, and 
$q$ ($\ep$) the momentum (polarization vector) of the photon.
In the last step we used $q=(p - p')$ and $q\ep=0\,$, where $p$ ($p'$)
 is the momentum
of the $b$- ($s$-) quark.
When calculating a given Feynman diagram, it is sufficient to work out
only the term proportional to $(p\ep)\,R$. After adding all the diagrams,
the full answer can be reconstructed by means of eq. (\ref{decomp}).
This reduces the number of diagrams; e.g., 
when calculating the $O(\a_s)$ corrections for $b \to s \gamma$
in the full theory,
"only" the graphs in Fig. 2 have a non-zero projection 
on the term $(p\ep)R$.\\ 
A similar projection for the process
$b \to s g$ (with obvious changes) can also be obtained.

\subsection{Method for calculating of the two-loop diagrams}
To extract $C_{71}$ and $C_{81}$ various one- and two- loop
diagrams have to be calculated in both versions (full/effective) 
of the theory. As the one-loop diagrams are straightforwardly
obtained by conventional techniques, we directly move to the two-loop
diagrams. 
When working out $b \to s \g$ and $b \to s g$ in the effective
theory at the matching scale $\mwt$, 
the only two-loop contributions leading to terms of order $\a_s$ 
are those associated with the operator
$O_2$. For the $b \to s \g$ case, these terms have been obtained in 
\cite{GHW}. We anticipate, that in the corresponding full theory 
calculation a term appears which can  be identified 
with the $O_2$ contribution in the effective theory. Consequently,
the $O_2$ contribution is not needed explicitly for extracting
$C_{71}$ and $C_{81}$.

Therefore, we directly discuss the calculation of  the two-loop 
contributions in the full theory.
In order to match dimension 6 operators, it is sufficient to 
extract the terms  of order $m_b \, \frac{m_b^2}{M^2} $ ($M=m_W,m_t$) 
from the full-theory
matrix elements for $b \to s \g$ and $b \to s g$
(term supressed by additional powers of 
$\mb/M$ correspond to higher 
dimensional operators in the effective theory). 
A systematic expansion of the matrix elements in inverse 
powers of 
$M$
can naturally be obtained by using the
well-known Heavy Mass Expansion 
(HME). 
In our context we use this HME only as a method for working out the
dimensionally regularized two-loop Feynman graphs
(and not to get directly renormalized quantities).
The theory of asymptotic expansions  of Feynman diagrams 
is already a textbook matter \cite{smir1}~\footnote{The idea 
of deriving operator product expansions using subtractions of leading 
asymptotics  goes back to Zimmermann \cite{Zim}. Later this idea was 
systematically developed within the BPHZ scheme \cite{AnZa}. The  simple 
explicit formulae for asymptotic expansion within dimensional
regularisation like (\ref{HMEbasic}) 
have been systematically derived in \cite{HMEsum}.}. Therefore, we 
only recall 
those properties of the HME, which are
of practical importance for our calculation  (for the mathematical 
foundations  of this method we refer to the literature \cite{HMEsum}):\\
\vspace*{0.1cm}

Suppose that all the masses of a given Feynman diagram $\Gamma$
can be divided into a
set of large \mbox{$\underline{M}=\{M_{1},M_{2}, \ldots$\}} and small
$\underline{m}=\{m_{1},m_{2}, \ldots\}$ masses  and assume that
all external momenta
$\underline{q}=\{q_{1},q_{2}, \ldots\}$ are small compared to the scale of
the large masses $\underline M$; then the statement is that  
the dimensionally regularised (unrenormalized) Feynman integral 
$F_{\Gamma}$ 
associated with the Feynman diagram
$\Gamma$ can we written as
\be
\label{HMEbasic}
F_{\Gamma} \stackrel{\underline{M} \to \infty}{\sim}
\sum_{\gamma} F_{\Gamma / \gamma} \circ
{\cal T}_{\underline{q}^{\gamma},\underline{m}^{\gamma}} 
F_{\gamma}(\underline{q}^{\gamma},\underline{m}^{\gamma},\underline{M})
\quad ,
\ee
where the sum is performed over all subgraphs $\gamma$ of $\Gamma$ which 
fulfill the following two conditions simultaneously:\\
$\bullet$ $\gamma$  contains all lines with heavy masses ($\underline{M}$) and \\
$\bullet$ $\gamma$ consists of connectivity components that are
one-particle-irreducible with respect to lines with small masses ($\underline{m}$).\\
\vspace*{0.1cm}

Here some clarifying remarks are in order:\\ 
$\bullet$ The operator ${\cal T}$ performs a Taylor expansion in the variables
$\underline{q}^{\gamma}$ and $\underline{m}^{\gamma}$,
where  $\underline{m}^{\gamma}$ denotes  the set of light masses in
$\gamma$ and $\underline{q}^{\gamma}$ denotes the set of all 
external momenta with respect to the subgraph $\gamma$; 
to be more specific, an external momentum with respect to the subgraph $\g$ 
can be an internal momentum with respect to the full graph $\Gamma$. 
 $F_{\Gamma / \gamma}$ denotes the Feynman integral corresponding 
to the reduced graph $\Gamma/ \gamma$. Note that the operator ${\cal T}$ is understood to 
 act directly on the integrand of the subgraph $\gamma$. 
The
decomposition of the original, say $l$-loop-diagram 
$\Gamma$
into the subdiagram $\gamma$ and 
the diagram $\Gamma/ \gamma$ is achieved in the corresponding Feynman 
integral by factorizing the 
product of scalar
propagators as $\Pi_{\Gamma} = \Pi_{\Gamma / \gamma} \Pi_{\gamma}$ 
such that
\be
\label{faktor}
F_{\Gamma / \gamma} \circ  {\cal T}_{\underline{q}^{\gamma},
\underline{m}^{\gamma}}       F_{\gamma} =
\int dk_{1} \cdots  dk_{l} \quad \Pi_{\Gamma / \gamma} 
\quad {\cal T}_{\underline{q}^{\gamma},
\underline{m}^{\gamma}}   \Pi_{\gamma}.
\ee
$\bullet$ The full graph $\G$ is always a subgraph contributing in the
sum $\sum_{\gamma}$.\\ 
$\bullet$
It is instructive to look at the
special case where 
all masses are large compared to the external momenta in a 
given diagram $\G$. In this case
only the full graph $\Gamma$  contributes to the sum $\sum_{\gamma}$
in (\ref{HMEbasic}).
The complete HME expansion reduces to a naive  
Taylor expansion in the external momenta of the  
integrand of the Feynman integral:
\be
\label{HME1}
F_{\Gamma} \stackrel{\underline{M} \to \infty}{\sim}
{\cal T}_{\underline{q}^{\Gamma}} \, 
F_{\Gamma}(\underline{q}^{\Gamma},\underline{M})
\ee
$\bullet$ The Taylor operator ${\cal T}$ introduces
additional spurious IR- or UV-divergences in the various terms of the 
sum $\sum_{\gamma}$, as we will see in an explicit example below.
It is a nontrivial property of the HME that all
these artificial
divergences cancel after making a summation over all possible
subgraphs $\gamma$. For our calculations  
this property  provides 
a nontrivial check for the individual diagrams,
as this cancellation has to happen diagram by diagram. \\

Now we illustrate this rather formal description for the diagram in Fig 1b,
for an internal top quark and denote it 
$D_{top}$. It belongs to the Set1 in Fig 2.  The $W$- and $\Phi$-exchange contributions are understood 
to be added.
The corresponding Feynman integral has the following form 
(the Dirac spinors  $\overline{u}(p')$ and $u(p)$ are amputated):
\begin{eqnarray}
\label{HME2}
D_{top} &=& X \, \exp(2 \g_E \e) \mu^{4 \e} (4\p)^{-2\e}
 \int \frac{d^dr}{(2\pi)^d} \int  \frac{d^dl}{(2\pi)^d} \, \times
  \\
 && \hspace{-1cm}
\frac{Dirac_{1t}}{(p-q+r)^2 - m_s^2} \, 
\frac{1}{r^2} \, \frac{Dirac_{2t}}{
\left[\left((l+r)^2 - m_t^2 \right) \, \left(l^2 - m_t^2 \right) \,
\left((l+q)^2 - m_t^2 \right) \, \right]} \,
 \frac{1}{(l+p-q)^2 - m_{W}^2} \nonumber
\end{eqnarray}
In (\ref{HME2}) the functions $Dirac_{1t}$ and $Dirac_{2t}$ are
the respective Dirac structures, whose 
explicit form is not important for explaining the 
principle steps of the expansion.
The constant
$X$ collects all the remaining constant factors like 
coupling constants and
CKM-factors. 

We find two subdiagrams $\gamma$ of $D_{top}$ 
which fulfill the two conditions 
given below eq. (\ref{HMEbasic}):
The first contribution of the HME 
corresponding to the subdiagram $\gamma_1$ 
shown in Fig. 1c is given
by
\begin{eqnarray}
\label{HME3}
D_{top}^{\bf 1} &=& X \, \exp(2 \g_E \e) \mu^{4 \e} (4\p)^{-2\e} 
\int \frac{d^dr}{(2\pi)^d} \int  \frac{d^dl}{(2\pi)^d} \, 
\times   \\
&& \hspace{-1.3cm}   
\frac{Dirac_{1t}}{(p-q+r)^2 - m_s^2} \, 
\frac{1}{r^2} \, {\cal T}_{r,p,q} \, \left( \, \frac{Dirac_{2t}}{
\left[\left((l+r)^2 - m_t^2 \right) \, \left(l^2 - m_t^2 \right) \,
\left((l+q)^2 - m_t^2 \right) \, \right]} \,
 \frac{1}{(l+p-q)^2 - m_{W}^2} \, \right) \quad . \nonumber 
\end{eqnarray}
The second contribution is the naive one, $\gamma_2 = D_{top}$ (see Fig. 1d):
\begin{eqnarray}
\label{HME4}
D_{top}^{\bf 2} &=& X \, \exp(2 \g_E \e) \mu^{4 \e} (4\p)^{-2\e}
 \int \frac{d^dr}{(2\pi)^d} \int  \frac{d^dl}{(2\pi)^d}   \,
\times \\
&& \hspace{-1.3cm}
{\cal T}_{p,q} \, \left( \, \frac{Dirac_{1t}}{(p-q+r)^2 - m_s^2} \, 
\frac{1}{r^2} \, \frac{Dirac_{2t}}{
\left[\left((l+r)^2 - m_t^2 \right) \, \left(l^2 - m_t^2 \right) \,
\left((l+q)^2 - m_t^2 \right) \, \right]} \,
 \frac{1}{(l+p-q)^2 - m_{W}^2} \right) \quad . \nonumber 
\end{eqnarray}
So we end up with
\be
\label{HME5}
D_{top}
\stackrel{\underline{M} \to \infty}{\sim}
D_{top}^{\bf 1} + D_{top}^{\bf 2} \quad .
\ee

The integrals are considerably simplified after the Taylor operation
${\cal T}$ and can be solved analytically after introducing 
Feynman parametrization. We mention that the Dirac algebra has been 
done with the algebraic program REDUCE \cite{REDUCE} and the
integrals have been done with the symbolic program MAPLE \cite{MAPLE}.

As mentioned above
we can discard terms of order $1/M^3$. 
Simple dimensional arguments tell us that we have to 
perform the Taylor operation ${\cal T}$ 
up to second order in the external
momenta $r$, $p$, $q$   in $D_{top}^{\bf 1}$
and also up to second order in $p$, $q$ in $D_{top}^{\bf 2}$. 
Restoring all the factors which we symbolized by $X$, 
and projecting on the term $(p\ep)\,R$,
we get
\be
\label{sets2}
D_{top} = \frac{4 \, i \, G_F \, \l_t}{\sqrt{2}} \, \frac{\a_s}{4\p}
\, C_F \, \frac{e}{16\pi^2} \, (-4 \, m_b) \, (p\ep)R \,
\left[ d_{top}^{1} + d_{top}^{2} \right] \quad .
\ee
The quantities $d_{top}^{1}$ and $d_{top}^{2}$ are given by 
($z=(m_t/m_W)^2$):

\begin{eqnarray}
\label{d1}
d_{top}^1 &=& -\frac{S}{\e} \, \left( 1 - 2 \, \ln (\mb/\mw) \, \e -
2 \, \ln (m_s/m_b)\, \e + 4 \, \ln (\m/\mw) \, \e \, \right) -
\frac{9z^2-8z+2}{36(z-1)^4} \, \ln^2 z \nonumber \\
&& + \frac{11z^4-14z^3+234z^2-180z+24}{108(z-1)^4} \, \ln z
- \frac{229z^3+15z^2+744z-538}{648(z-1)^3}  
\end{eqnarray}

\begin{eqnarray}
\label{d2}
d_{top}^2 &=& + \frac{S}{\eir} \, \left( 1 + 4 \, \ln (\m/\mw) \, \e \, 
\right) + \frac{-108z \, \ln z +60z^4-258z^3+468z^2-294z+24}{108(z-1)^4 \, \e} 
\, \left( \frac{\m}{\mw} \, \right)^{4\e} \nonumber \\
&& + \frac{9z^2+10z+2}{12(z-1)^4} \, \ln^2 z
-\frac{142z^4-538z^3+753z^2-218z+26}{108(z-1)^4} \, \ln z \nonumber \\
&& - \frac{z^4-40z^3+27z^2-10z-2}{18(z-1)^4} \, \Li 
- \frac{67z^3+2343z^2-2766z+230}{648(z-1)^3} \quad ,
\end{eqnarray}
where the function $S$ is
\be
\label{sfun}
S = \frac{(-54z^2+48z-12)\, \ln z + 11z^4-14z^3+27z^2-38z+14}{108(z-1)^4}
\quad .
\ee

The $1/\e$ poles in $d^1_{top}$ correspond to
spurious ultraviolet singularities produced in the $r$-integration
after expanding the subdiagram $\g_1$.
The $1/\eir$ poles in $d^2_{top}$ 
on the other hand arise due to the worsened
infrared behaviour induced when expanding the $s$-quark propagator.
As we explicitly
see, these artifical singularites cancel when adding 
$d^1_{top}$ and $d^2_{top}$.

We now discuss the corresponding diagram $D_{charm}$ where the 
internal top quark is
replaced by the (light) charm quark.
The quantities 
$D_{charm}^{\bf 1}$ and $D_{charm}^{\bf 2}$ 
corresponding to the 
subdiagrams $\gamma_1$ and $\gamma_2$ (see Fig. 1c,d)
are given by the 
analogous formulae (\ref{HME3}) and (\ref{HME4}), where $m_t$ is replaced
by $m_c$ and the Taylor operator ${\cal T}_{r,p,q}$ in (\ref{HME3}) is
replaced by ${\cal T}_{r,p,q,m_c}$ and
${\cal T}_{p,q}$ in  (\ref{HME4}) by   ${\cal T}_{p,q,m_c}$.
As we are discarding terms of order $1/M^3$, it turns out that
only the zeroth order term in the $m_c$ expansion has to be retained;
this amouts to putting $m_c=0$ in $D_{charm}^{\bf 1}$ and 
$D_{charm}^{\bf 2}$. 

Moreover, in the charm-case there is a third contribution 
to the HME which
corresponds to the subdiagram $\gamma_3$ in Fig. 1e.
The latter
consists of the $W/\Phi$-line only. 
As we neglect terms of order $1/M^3$, the Taylor expansion
of the corresponding Feynman integral just amounts to 
replace the $W$ and $\Phi$ propagator
by $i/m_W^2$ and $-i/m_W^2$, respectively. 
As the Feynman integral of the $\Phi$ diagram
has an additional factor of order $(m_c \, m_b)/m_W^2$ from the
Yukawa couplings, only the four Fermi version of the $W$ 
exchange diagram effectively contributes to
$D_{charm}^{\bf 3}$. Stated differently, $D_{charm}^{\bf 3}$
is directly related to the $O_2$ contribution in the effective theory.
Of course, this is not suprising when keeping in mind 
how the effective Hamiltonian is constructed. 

To summarize, $D_{charm}$ is given by 
\be
\label{HME7}
D_{charm}
\stackrel{\underline{M} \to \infty}{\sim}
D_{charm}^{\bf 1} + D_{charm}^{\bf 2} + D_{charm}^{\bf 3} \quad .
\ee

\subsection{The matching to leading-log precision}
To establish some lowest order matching
functions which are frequently used in the following sections and
in order to explain  an important subtelty in the NLL matching
calculation, we recall the results 
of the LL matching: 
In the full theory  
the lowest order matrix elements $M_0$ for $b \to s \g$
and $b \to s g$ 
are obtained by expanding the diagrams
shown in Fig. 1a up to second order in the external momenta.
The results read in $d=4-2\e$ dimensions
\be
\label{lowestfull}
M_0(b \to s \g) = \frac{4 \, i \, G_F \, \l_t}{\sqrt{2}} \,
K_{70} \, \bra s \g|O_7|b\ket \quad , \quad
M_0(b \to s g) =  \frac{4 \, i \, G_F \, \l_t}{\sqrt{2}}  \,
K_{80} \, \bra s g|O_8|b\ket \quad ,
\ee
where the functions $K_{70}$ and $K_{80}$ have an expansion in $\e$ of the form
\be
\label{ceps}
K_{70} = K_{700} + \e \, K_{701} + \e^2 \, K_{702} + \ldots \quad , \quad
K_{80} = K_{800} + \e \, K_{801} + \e^2 \, K_{802} + \ldots \quad .
\ee 
On the other hand, the lowest order result $\hat{M}_0$ in the effective
theory reads (also in $d=4-2\e$ dimensions)
\be
\label{lowesteff}
\hat{M}_0(b \to s \g) = C_{70} \, \bra s \g|O_7|b\ket \quad , \quad
\hat{M}_0(b \to s g) = C_{80} \, \bra s g|O_8|b\ket \quad .
\ee
As the matching is understood to be done in 4 dimensions,
we get the connections
\be
\label{match0}
C_{70} = K_{700} \quad , \quad 
C_{80} = K_{800} \quad .
\ee
Therefore, in $d$ dimensions
$M_0$ and $\hat{M}_0$ differ by terms of order $\e$.
This detail becomes an important subtelty when going to higher 
loop orders; we best explain this by means of an example:
one type of order $\a_s$ corrections is given by
multiplying the lowest order result by ultraviolet singular 
$\sqrt{Z_2}$ factors
which are the same for both versions of the theory.
In the full theory,
this leads to finite terms proportional to $K_{701}$ (and $K_{801}$);
the corresponding terms in the effective theory are not generated.

When working out
the two-loop  integrals corresponding to the diagrams in
Figs. 2 and 3 in the full theory
for $b \to s \g$ (or $b \to s g$), there are contributions in which the 
dimensionally regularized lowest order
result, taken up to first or second order in $\e$, factorizes. 
As we will see later, the infrared singularity stucture is precisely
of this form.
As we will use the explicit expressions for the 
Inami-Lim \cite{Inami} functions $K_{700}$,
$K_{800}$, $K_{701}$ and $K_{801}$ at several places, we list them here.
Using $z=(\mt/\mw)^2$, they read
\bea
\label{k700}
K_{700}=C_{70} &=& \frac{z\,\left[6z(3z-2) \, \ln z -(z-1) \, 
(8z^2+5z-7)\right]}{24(z-1)^4} \\
\label{k800}
K_{800}=C_{80} &=& - \frac{z \left[ 6z \, \ln z + (z-1) \, (z^2-5z-2) \,
\right]}{8(z-1)^4} 
\eea

\bea
K_{701} &=& -\frac{z\, \left[ \, 18z(3z-2)\, \ln^2 z + (44z^3-314z^2+324z-96) \, \ln z + 56z^3-35z^2-56z+35 \, \right]}{144(z-1)^4}
\nonumber \\
&& + 2 \, K_{700} \, \ln \, (\m/\mw)
\eea

\bea
\label{ceps2}
K_{801} &=& -\frac{z\, \left[ -18z \, \ln^2 z + 
(10z^3-28z^2+108z-48) \, \ln z + 25z^3-118z^2+119z-26 \, \right]}{48(z-1)^4}
\nonumber \\
&& + 2 \, K_{800} \, \ln \, (\m/\mw)
\eea

\section{$b \to s \gamma$ in the full theory}
\label{sec:fullbsgamma}
\setcounter{equation}{0}
In section 3.1 we present the results for the dimensionally
regularized matrix element $M$ for $b \to s \g$ in the full theory.
In section 3.2 we discuss the various counterterm contributions. 
\subsection{Two-loop Feynman diagrams}
As in eq. (\ref{expo}), we write 
the $b\to s \g$ matrix element $M$ in the
form $M= M_0 + \frac{\a_s}{4\pi} \, M_1$. When using the "reduction
technique" described in section 2.3, the complete list of  two-loop
diagrams contributing to $M_1$ is given in Fig. 2, where
the cross stands for the possible locations where the photon can 
be emitted. 
Note that diagram 5b in Fig. 2 does not contribute in the
limit $m_s=0$.
We write the result for $M_1$ in the form
\be
\label{sets}
\frac{\a_s}{4\pi} \, M_1 = \, V \,
\left[ R_t^{1+2} - R_c^{1+2} - R_c^3 \right]
\quad ,
\ee  
where $V$ is an abbreviation for the often occurring quantity
\be
\label{v}
V=\frac{4 \,i\, G_F \, \l_t}{\sqrt{2}} 
\, \frac{\a_s}{4 \p} \, C_F \,
\bra s \gamma |O_7 | b \ket_{tree} \quad ; \quad C_F=\frac{4}{3} \quad .
\ee
In eq. (\ref{sets}) $R_t^{1+2}$ ($R_c^{1+2}$) denotes the sum of the
first and second contribution in the Heavy Mass Expansion (HME) 
(see section 2.4) 
of the dimensionally regularized (unrenormalized) Feynman integrals 
for internal top (charm) quark;
$R_c^3$ is the third contribution  in the HME,
which has to be considered only for the light internal quarks,
which in our present case is the charm quark ($\lambda_u=0$). 
According to the HME, $R_c^3$ is
obtained by working out the charm loops using the four-Fermi
approximation of the $W$-propagator. Stated differently, $R_c^3$
is directly related to the order $\a_s$ contribution of
matrix element of the operator $O_2$, 
provided the latter is evaluated in the NDR scheme;
more precisely,
\be
\label{ident}
R_c^3 = -\hat{R}_2 \quad
\ee
where $\hat{R}_2$ is the quantity defined through the equation  
\be
\label{rc3o2}
 \bra s \g |O_2| b \ket =  \frac{\a_s}{4\pi} \, C_F \, 
\bra s \gamma |O_7 | b \ket_{tree} \, \hat{R}_2 \quad .
\ee   
As the same contribution is also present in the effective theory,
we will not need to know
$R_c^3$ explicitly 
\footnote{The reader who whishes to see the explicit form
for $\hat{R}_2$ is referred to eq. (2.35) in ref. \cite{GHW}.}
in order to extract 
the order $\a_s$ corrections in the Wilson
coefficient $C_{71}(\mwt)$.
Making use of the various $K-$functions given in 
(\ref{k700})--(\ref{ceps2})
and denoting $r=(m_s/m_b)^2$, we now give the dimensionally regularized 
expression for
$R^{1+2} \equiv R_t^{1+2} - R_c^{1+2}$. 
\bea
\label{r12tc}
R^{1+2}  &=& - (K_{700} + \e \, K_{701})
\, \frac{\left( \frac{\m}{\mw} \right)^{2 \e}}{\eir} \, \ln r +  
g_1 \, \frac{\left( \frac{\m}{\mw} \right)^{4 \e}}{\e}
+ \frac{1}{2} \, K_{700} \, \ln^2 r \nonumber \\
&&
+ 2 \, K_{700} \, \ln r \ln (\mb/\mw) - 2 \, K_{700} \, \ln r + g_2 \ln
(\mb/\mw) + g_3 \quad .   
\eea 
The first term in eq. (\ref{r12tc}) 
is due to infrared singularities in the on-shell 
$ b \to s \gamma$ amplitude as suggested by the notation
\footnote{We could separate ultraviolet and infrared poles in our calculation.
In the follwing, $1/\eir$ ($1/\e$) stands for infrared (ultraviolet) poles. } 
$1/\eir$. This term is entirely due to those diagrams in set 3 of Fig. 2
where the photon is radiated from  
the internal quark or the $W$ (or $\Phi$) boson.
The quantities 
$g_1$, $g_2$ and $g_3$ in eq. (\ref{r12tc}) can 
be written as ($z=(m_t/m_W)^2$, $\mbox{Li}(x)=-\int_0^x \,
\frac{dt}{t} \, \ln (1-t)$ )
\be
\label{g1}
g_1 = \frac{(-324z^4-450z^3+270z^2+72z)\, \ln z+
112z^5+244z^4+55z^3-931z^2+593z-73}{72(z-1)^5} - \frac{35}{216} 
\ee

\be
\label{g2}
g_2 = - \frac{(-216 z^3+162z^2-72z) \, \ln z +
44z^4+154z^3-393z^2+274z-79}{36(z-1)^4} - \frac{7}{4} 
\ee

\bea
\label{g3}
g_3 &=&  - \frac{z(8z^3+61z^2-40z+4) \,}{6(z-1)^4} \, \Li  +
\frac{2}{3} \, i \, \pi \, K_{800} +\frac{2}{27} \, \pi^2
+ \frac{3155}{1296} \nonumber \\
&& - \left[(-4860z^4-18954z^3+11502z^2+648z)\, \ln^2 z \right. \nonumber \\
&& + (3240z^5+16956 z^4+37638z^3-56586z^2+20688z-2496-216 \, \pi^2(z^3-z^2))
 \, \ln z \nonumber \\
&& +(-1442 z^5-55910z^4+109651z^3-69271z^2+20999z-4027) \nonumber \\
&& \left. + (60z^5-228z^4+636z^3-924z^2+552z-96) \, \pi^2 \, \right]
\, /(1296(z-1)^5)
\eea

\subsection{Counterterms}
The counterterms relevant for calculating on-shell matrix elements
are generated by expressing the bare parameters in the original
Lagrangian in terms of the renormalized quantites. Working up to
order $\a_s$, the only parameters which need renormalization in the
present situation are the $t$-quark mass and the $b$- quark mass
(in principle also the $s$- quark mass if we did not work in the
limit $m_s=0$). Using on-shell renormalization for the
external $b$-quark mass 
and $\overline{MS}$ renormalization for the (internal)
top quark mass, the connection between the bare and renormalized masses
reads
\bea
\label{masscount}
m_{t,bare} &=& m_t - \delta m_t \quad , \quad
\frac{\delta m_t}{m_t} = \frac{\a_s}{4 \pi} \, C_F \,
 \frac{3}{\e} \nonumber \\
m_{b,bare} &=& m_b - \delta m_b \quad , \quad
\frac{\delta m_b}{m_b} = \frac{\a_s}{4 \pi} \, C_F \, \left( 
\frac{3}{\e} + 6 \, \ln (\m/\mb) +4 \, \right)
\eea
Note, these mass shifts not only shift the mass terms like 
$m_t \, \bar{t} \, t$, 
but also the Yukawa terms like 
$\sim g \, \bar{b} \, (m_b L - m_t \,R \,) t \, \Phi^-$,
where $\Phi^-$ is the unphysical charged Higgs
field which appears in covariant gauges.
These counterterms, induced by the shifts $\delta m_t$ and 
$\delta m_b$, generate corrections for the $b \to s \g$
matrix element, which we denote by $\delta M_b$ and $\delta M_t$, 
respectively.
Writing $\delta M_{f} = V \, \delta R_{f}$ ($f=t,b$) with $V$ given
in eq. (\ref{v}), we get 
\bea
\label{deltarb}
\delta R_b &=&  \left\{
\left((6z-8) \, \ln z - 7 z^2 +16z - 9 \right) \, \left(
\frac{2}{\e} + 4 \, \ln(\m/\mb) +8/3 \, \right) \right. \nonumber \\
&& \hspace{-1.0cm} \left. +
(-6z+8)  \, \ln^2 z + (20z^2-26z) \, \ln z
-19z^2+44z-25 \, \right\} \, 
\left( \frac{\m}{\mw} \, \right)^{2\e} \, \frac{z}{16(z-1)^3} 
\eea
while $\delta R_t$ is given by
\begin{eqnarray}
\label{deltart}
\delta R_t &=&  
\left\{
\frac{6}{\e} \, \left(
(18z^3+30z^2-24z) \, \ln z -47z^3+63z^2-9z-7 \, \right) 
+ 18z \, (-3z^2-5z+4) \, \ln^2 z  \right. \nonumber \\ 
&& \hspace{-1.0cm} \left. +
(246z^3+114z^2-288z+96)  \, \ln z +
44z^4-547z^3+855z^2-413z+61 \, \right\} \, \times \nonumber \\
&& \left( \frac{\m}{\mw} \, \right)^{2\e} \, 
\frac{z}{24 \, (z-1)^5} 
\end{eqnarray}
When writing down the expression for $\delta R_b$, we should mention,
that we did not include the insertion of $\delta m_b \, \bar{b} \, b$
in the external $b$-quark leg. This is quite in analogy of omitting
self-energy diagrams for the external legs.
Such corrections on the external legs are taken into account by multiplying
the amputated diagram with the factor 
$\sqrt{Z_{2,b} \, Z_{2,s}}$, where $Z_{2,b}$ and $Z_{2,s}$
are the residues taken at the (physical) pole position of the regularized
$b$- and $s$- quark two point functions, respectively.
Making use of the expression (in Feynman gauge) 
\be
\label{deltarz2}
Z_2(m) = 1 - \frac{\a_s}{4\pi} \, C_F \, \left( \frac{\m}{m} \right)^{2\e}
\, \left[ \, \frac{1}{\e} + \frac{2}{\eir} + 4 \, \right] \quad ,
\ee
the counterterm $\delta M_{Z_2}$ induced by the $Z_2$- factors
of the external quark fields reads (again writing 
$\delta M_{Z_2} = V \, \delta R_{Z_2}$)
\bea
\label{deltaz}
\delta R_{Z_2} &=& - 
\left(
\frac{\m}{\mw} \right)^{2\e} \, 
\left\{
\frac{2}{\eir} \, \left( K_{700} + \e \, K_{701} \right) \, 
+
\frac{1}{\e} \, \left( K_{700} + \e \, K_{701} \right) \, 
\right. \nonumber \\
&& \left. + \left(4- 6 \, \ln (\mb/\mw) - \frac{3}{2} \, \ln r \, \right) \,
K_{700} \, \right\} \, 
 \quad .
\eea

\section{$b \to s \gamma$ in the effective theory}
\label{sec:effbsgamma}
\setcounter{equation}{0}
As in the full theory, we first discuss the matrix elements
for $b \to s \g$ of the operators in basis (\ref{operators}).
In section 4.2 we list the various counterterm contributions.
\subsection{Regularized Feynman diagrams}
We write the matrix element $\hat{M}$ for $b \to s \g$ as a sum
of the contributions due to the operators $O_i$ in the effective
Hamiltonian, i.e.,
\be
\hat{M} = \sum_{i=1}^8 \hat{M}^i \quad ; \quad 
\hat{M}^i = \frac{4 i G_F \l_t}{\sqrt{2}} \, C_i \,
\bra s \g|O_i|b \ket \quad . 
\ee
To facilitate later the comparison between the results in the two
versions of the theory (full vs. effective), we write $\hat{M}^i=
\hat{M}_0^i + \frac{\a_s}{4\p} \, \hat{M}_1^i$
 and cast the term proportional to $\a_s$
in the form 
\be
\label{effres}
\frac{\a_s}{4\p} \,
\hat{M}_1^i = V \, \hat{R}_i \quad , 
\ee
where $V$ is given in eq. (\ref{v}).

We first discuss the contributions of the four-Fermi operators
$O_1$--$O_6$.
As the Wilson coefficients of  $O_1$,
$O_3$, $O_4$, $O_5$ and $O_6$ start at order $\a_s^1$, we 
only have to take into account 
their order $\a_s^0$ (one-loop) matrix elements;
 it is well-known
that in the NDR scheme
only $O_5$ and $O_6$ have a non-vanishing one-loop 
matrix element for $b \to s \g$. 
Making use of the Wilson coefficients (see \cite{Buras})
\be
\label{c5c6}
C_5(\m) = \frac{\a_s(\m)}{4\pi} \, C_F \,
 \left[-\frac{1}{6} \, \ln \frac{\m}{\mw} - 
\frac{1}{8} \tilde{E} \right] \quad , \quad
C_6(\m) = \frac{\a_s(\m)}{4\pi} \, C_F \,
\left[\frac{1}{2} \, \ln \frac{\m}{\mw} + 
\frac{3}{8} \tilde{E} \right] \quad , \quad
\ee
$\hat{R_5}$ and $\hat{R}_6$ are readily obtained
\be
\label{rhat5}
\hat{R}_5 = - \frac{1}{3} \, 
\left[ \, -\frac{1}{6} \ln \frac{\m}{\mw} -
\frac{1}{8} \, \tilde{E} \, \right] \quad ,   \quad
\hat{R}_6 = -   
\left[ \, \frac{1}{2} \ln \frac{\m}{\mw} +
\frac{3}{8} \, \tilde{E} \, \right] \quad ,
\ee
with 
\be
\label{etilde}
\tilde{E} = -\frac{2}{3} \ln z + \frac{z^2(15-16z+4z^2)}{6(1-z)^4} \,
\ln z + \frac{z(18-11z-z^2)}{12(1-z)^3} - \frac{2}{3} \quad .
\ee
On the other hand,
the Wilson coefficient of the operator $O_2$
starts at order $\a_s^0$. Consequently,
we have to take in principle one- and two-loop
matrix elements of this operator. In practice, however, 
the order $\a_s^0$ (one-loop)
matrix element of $O_2$ vanishes and therefore only the contribution of the
order $\a_s^1$  (two-loop) matrix element remains:
\be
\label{rhat2}
\hat{R}_2 \quad .
\ee
As this contribution also occurs in the full theory result in section 3.1
(see eqs. (\ref{sets}) and (\ref{ident})),
the explicit expression for the r.h.s. of eq. (\ref{rhat2}) is not
needed for the extraction of $C_{71}$.

The order $\a_s$ contribution of the matrix elements of 
the dipole operator $O_7$ 
(see Figs. 4a,b) yields 
\be
\label{rhat7}
\hat{R}_7 = \frac{3}{4} \, C_{71} - C_{70} 
\, \frac{\left( \frac{\m}{\mw} \right)^{2 \e}}{\eir} \, \ln r +
\frac{C_{70}}{2} \, \ln^2 r 
+ 2 C_{70} \, \ln r \, \ln (\mb/\mw)
- 2 \, C_{70} \, \ln r  \quad .
\ee
The first term on the r.h.s. of eq. (\ref{rhat7}) comes from
the tree-level matrix element in Fig 4a,  
being multiplied with the
the order $\a_s$ part (i.e. $C_{71}$) of the Wilson coefficient
$C_7$. The remaining terms are due to the one-loop graph in Fig. 4b.  
Note that $C_{71}$ is the quantity we ultimately
wish to extract.
Finally, the diagrams of $O_8$ are depicted in Figs. 4c,d;
its contribution is \cite{GHW}
\be
\label{rhat8}
\hat{R}_8 = - \frac{C_{80}}{9} \, \left[-\frac{12}{\e} - 33 + 2 \pi^2 
+ 24 \ln(\mb/\m) - 6 \, i \, \pi \, \right] \quad .
\ee 

\subsection{Counterterms}
As the operators mix under renormalization, we have to consider
counterterm contributions induced by operators of the form
$C_i \, \delta Z_{ij} \, O_j$.
We denote their contributions to $b \to s \g$ by
\be
\delta \hat{M}_{ij} 
= \frac{4 \, i \, G_F \, \l_t}{\sqrt{2}} \, \bra s \g |C_i \,
\delta Z_{ij} \, O_j|b
\ket \quad .
\ee 
The non-vanishing matrix elements read 
(using $\delta \hat{M}_{ij} = V \, \delta \hat{R}_{ij}$)
\be
\label{m25}
\delta \hat{R}_{25} = 
\frac{1}{36} \, \frac{1}{\e} 
\, \left( \frac{\m}{\mb} \, \right)^{2\e} 
\ , \
\delta \hat{R}_{26} = -\frac{1}{4} \, \frac{1}{\e} 
\, \left( \frac{\m}{\mb} \, \right)^{2\e} \ , \
\delta \hat{R}_{27} = \frac{29}{27}  \, \frac{1}{\e} \ , \
\delta \hat{R}_{77} = \, \frac{4}{\e} \, C_{70} \ , \ 
\delta \hat{R}_{87} = - \, \frac{4}{3\e} \, C_{80} \ ,  
\ee 
where we made use of the renormalization constants \cite{counterterm}
\be
\label{renconst}
\left( \, \delta Z_{25}, \delta Z_{26}, \delta Z_{27}, \delta Z_{77},
       \delta Z_{87} \, \right) = \frac{\a_s}{4\p} \, C_F \, \left( \,  
- \frac{1}{12 \e}, \  \frac{1}{4 \e}, \ 
\frac{29}{27 \e}, \  \frac{4}{\e}, \   -\frac{4}{3\e} \, \right) \quad .  
\ee 
It is well-known that the renormalization of the four-Fermi
operators requires the introduction of counterterms proportional
to evanescent operators \cite{Burasweiss}. 
Calculating $b \to s \gamma$ up to order
$\a_s$, there are potential counterterm contributions involving
evanescent operators needed to renormalize $O_2$. As the initial
conditions for the four-Fermi operators (which we partially used in section
4.1) depend on the actual choice of the evanescent operators,
we have to use the same set when calculating their effect of $b \to s\g$.
We consistently take both, the initial conditions of the four-Fermi
operators and the set of evanescent operators from refs. \cite{Buras,
Burasweiss,Herrlich1,Herrlich2}.
The only potentially relevant matrix element of evanescent operators 
contributing $b \to s \gamma$ is
\be
\label{evform}
\bra s \g| \frac{1}{\e} \, E_1[O_2] \, |b \ket \quad , 
\ee
where the evanescent operator $E_1[O_2]$ is of the form 
\bea
\label{oev}
E_1[O_2] &=& \left[ \bar{s}_{\a_1}  \g_\m \g_\n \g_\eta  L  c_{\a_2}
 \, \bar{c}_{\a_3}  \g^\eta \g^\n \g^\m  L  b_{\a_4} -
(4+a_1 \e) \,
\bar{s}_{\a_1}  \g_\m  L  c_{\a_2}
 \, \bar{c}_{\a_3}  \g^\m \, L \, b_{\a_4} \, \right] \, 
K_{\a_1 \a_2 \a_3 \a_4} \nonumber \\
K_{\a_1 \a_2 \a_3 \a_4} &=&
\frac{1}{2}  \delta_{\a_1 \a_3}  \delta_{\a_2 \a_4}
-  \frac{1}{6} \delta_{\a_1 \a_2}  \delta_{\a_3 \a_4}
\quad .   
\eea
However, as these matrix elements are identically 
zero (in $d$ dimensions), there are no contributions
from counterterms proportional to evanescent operators.

Besides the counterterms induced by operator mixing, we also
have to renormalize the $b$-quark mass which explicitly appears in the
operator $O_7$ and in addition we have to multiply the lowest order 
matrix element by the factor $\sqrt{Z_2(m_b) \, Z_2(m_s)}$, 
quite in analogy to the calculation in the
full theory. 
The counterterm due to the $b$-quark mass renormalization
$\delta \hat{M}_b = V \, \delta \hat{R}_b$
yields 
\be
\label{deltahatrb}
\delta \hat{R}_b = - 
\left[ \, \frac{3}{\e} + 6 \, \ln (\m/\mb) \, +4 \,
\right] \, C_{70}  \quad ,
\ee
when using the on-shell definition for the $b$-quark mass, while
the counterterm $\delta \hat{M}_{Z_2}= V \, \delta \hat{R}_{Z_2}$ 
is given by
\be
\label{deltahatz2}
\delta \hat{R}_{Z_2} = - \left( \frac{\m}{\mw} \right)^{2\e} \, 
\left\{\frac{2}{\eir} \, C_{70} \, 
+
\frac{1}{\e} \, C_{70}  \, 
+ \left(4 - 6 \, \ln (\mb/\mw) - \frac{3}{2} \, \ln r \, \right) \,
C_{70} \, \right\}  \quad .
\ee

\section{Extraction of $C_{71}(\mwt)$}
\setcounter{equation}{0}
To summarize section 3, the order $\a_s$ part $M_1^{ren}$
of the renormalized matrix element 
for $b \to s \g$ in the full theory reads
\be
\label{resrenfull}
\frac{\a_s}{4\p} \, M_1^{ren} = V \, \left[
R^{1+2} + \hat{R}_2 + \delta R_b + \delta R_t + \delta R_{Z_2} \,
\right] \quad ,
\ee
where the quantities in the bracket on the r.h.s. of eq. (\ref{resrenfull})
are given in eqs. (\ref{r12tc}), (\ref{ident}), (\ref{deltarb}),
(\ref{deltart}) and (\ref{deltaz}), respectively; 
the prefactor $V$ is given in eq. (\ref{v}).

The corresponding renormalized matrix element $\hat{M}_1^{ren}$
in the effective theory
can be obtained from the information in section 4; $\hat{M}_1^{ren}$
reads
\be
\label{resreneff}
\frac{\a_s}{4\p} \, \hat{M}_1^{ren} = V \, \left[\hat{R}_2 + \hat{R}_5 +
\hat{R}_6 + \hat{R}_7 + \hat{R}_8 +
\delta \hat{R}_{25} + \delta \hat{R}_{26} 
+\delta \hat{R}_{27} + \delta \hat{R}_{77} + \delta \hat{R}_{87} 
+ \delta \hat{R}_b  + \delta \hat{R}_{Z_2} \,
\right] \quad ,
\ee
where the various quantities in the bracket are given in eqs.
(\ref{rhat2}), (\ref{rhat5}), (\ref{rhat7}), (\ref{rhat8}),
(\ref{m25}), (\ref{deltahatrb}) and (\ref{deltahatz2}). 

Before we are able to correctly extract $C_{71}$,
a remark concerning 
the infrared structure is  in order. 
We split $M_1^{ren}$ into a infrared singular and an infrared finite
piece, i.e., 
\be
\label{infrafull}
M_1^{ren}= M_{1,ir}^{ren} + M_{1,fin}^{ren}.
\ee
As this splitting is not unique (concerning the finite terms), we
define the singular part to be
\be
\label{fullirsing}
M_{1,ir}^{ren} = 
- (\, K_{700} + \e \, K_{701} \, ) \, 
\frac{\left( \frac{\m}{\mw} \right)^{2\e}}{\eir} \, \ln r 
- 2\, (\, K_{700} + \e \, K_{701} \, ) \, 
\frac{\left( \frac{\m}{\mw} \right)^{2\e}}{\eir}  \quad ,
\ee
where the first and second term on the r.h.s. are due to the two-loop
diagrams (\ref{r12tc}) and the counterterms 
(\ref{deltaz}), respectively.
We do now an analogous splitting for the renormalized matrix element
in the effective theory, i.e.,
\be
\label{infraeff} 
\hat{M}_{1}^{ren} =  \hat{M}_{1,ir}^{ren} + \hat{M}_{1,fin}^{ren} \quad ,  
\ee
with
\be
\label{effirsing}
\hat{M}_{1,ir}^{ren} = - C_{70}  \, 
\frac{\left( \frac{\m}{\mw} \right)^{2\e}}{\eir} \, \ln r 
- 2\, C_{70} \, 
\frac{\left( \frac{\m}{\mw} \right)^{2\e}}{\eir}  \quad .
\ee

As the matching has to be done in four dimensions, we 
cannot - strictly speaking - 
use the process $b \to s \gamma$ 
to do the matching, because of the infrared singularities.
To cancel these singularities, we
have to include 
the gluon Bremstrahlung process $b \to s \g g$ 
in both versions of the theory.
In the effective theory,
the process has been worked out in \cite{AG91,AG95,Pott} (but the explicit
result is not important here); the result in the
full theory is obtained from the effective theory result by
replacing $C_{70}$ by $K_{700} + \e K_{701}$. 
The correct physical matching condition  consists in requiring the
infrared finite quantity
$\Gamma = \Gamma (b \to s \g) + \Gamma(b \to s \g g; E_\g \ge
E_\g^{min})$
to be equal in both versions of the theory. Due to the specific
form of eqs. (\ref{infrafull}) -- (\ref{effirsing}) and due to the
specific difference in the bremsstrahlung contribution, it follows
that the physical matching condition implies
\be
\label{matchphys}
M_{1,fin}^{ren} = \hat{M}_{1,fin}^{ren} \quad .
\ee 

The extraction of $C_{71}$ is now straightforward. 
In summary: Writing the Wilson coefficient $C_7(\mwt)$  
at the matching scale $\mwt$ in the form
\be
\label{resc7}
C_7(\mwt) = C_{70}(\mwt) + \frac{\a_s}{4\p} \, C_{71}(\mwt) \quad ,
\ee
we obtain (in the naive dimensional regularization scheme)
\bea
\label{resc71}
C_{71}(\mwt) &=& -\frac{2z\,(8 z^3 + 61z^2 - 40z+4)}{9(z-1)^4} \, \Li
+ \frac{2z^2\,(3z^2+23z-14)}{3(z-1)^5} \, \ln^2 z \nonumber \\
&& - \frac{2\,( 51 z^5+294z^4+1158z^3-1697z^2+742z-116)}{81(z-1)^5} \, \ln z
\nonumber \\
&& + \frac{1520z^4 + 12961 z^3 -12126z^2 + 3409z-580}{486(z-1)^4} 
\nonumber \\
&& - \frac{4z^2\,(3z^2+23z-14)}{3(z-1)^5} \, \ln z \, \ln (\mwt/\mw)
\nonumber \\
&& + \frac{2\,(106z^4+287z^3+1230z^2-1207z+232)}{81(z-1)^4} \, \ln (\mwt/\mw)
\quad .
\eea
Here, $z=(m_t(\mwt)/m_W)^2$, where $m_t(\mwt)$ is the 
$\overline{MS}$ top quark mass at the renormalization scale $\mwt$.
The lowest order function $C_{70}$ is given in eq. (\ref{k700}).

Taking into account that the result of Adel and Yao \cite{Adel}
is given in the so-called $R^*$ renormalization scheme,
we got the same result for $C_{71}(\mwt)$. 

\section{$b \to s g$ in the full theory}
\setcounter{equation}{0}
As in the $b \to s \gamma$ case we first give the
results for the two-loop diagrams and then move to the counterterm
contributions.
\subsection{Two-loop Feynman diagrams}
We again write the $b \to s g$ matrix element $M$ in the form
$M=M_0+\frac{\a_s}{4\p}\,M_1$. Using the 
"reduction technique" described in section
2.3, the complete set 
of two-loop Feynman graphs is given by the abelian diagrams in Fig. 2
and by  the non-abelian diagrams in Fig. 3, which involve the triple
gluon coupling. The crosses in Fig. 2 and Fig. 3 show the possible
locations from where the gluon can be emitted. 
Of course the graphs with a cross at the $W$ line in Fig. 2
have to be omitted.
Working in the limit
$m_s=0$, diagram 5b in Fig. 2 vanishes.
It is convenient to write $M_1$ in the form 
\be
\label{setsgluon}
\frac{\a_s}{4\p} \, M_1 = W
\left[ Q_t^{1+2} - Q_c^{1+2} - Q_c^3 \right]
\quad ,
\ee
where the quantity $W$ is defined as
\be  
\label{w}
W = \, \frac{4 \, i \, G_F \, \l_t}{\sqrt{2}} 
\, \frac{\a_s}{4 \p} \,
\bra s \gamma |O_8 | b \ket_{tree} \quad .
\ee
In eq. (\ref{setsgluon})
$Q_f^{1+2}$  denotes the sum of the
 first and second contribution in the 
Heavy Mass Expansion for an internal quark of flavor $f$ ($f=t,c$);
$Q_c^3$ is the third  contribution in this expansion,
which only has to be considered for the light internal quarks. 
Like $R_c^3$ in eq. (\ref{sets}) 
of section 3.1, $Q_c^3$ is just
\be
\label{identg}
Q_c^3 = -\hat{Q}_2 \quad ,
\ee
where $\hat{Q}_2$ is the quantity  
defined through the relation
\be
\bra s g |O_2|b \ket =  \frac{\a_s}{4\pi} \, 
\bra sg|O_8|b\ket_{tree} \, \hat{Q}_2 \quad . 
\ee
As exactly the same term also appears in the effective theory,
$Q^3_c$
 drops out when extracting the $O(\a_s)$ correction to 
the Wilson coefficent $C_8$. 

The dimensionally regularized expressions for
$Q^{1+2} \equiv Q_t^{1+2} - Q_c^{1+2}$ can be written in the form
\bea
\label{q12tc}
Q^{1+2} 
&=& \frac{1}{6} \, (K_{800} + \e \, K_{801})
\, \frac{\left( \frac{\m}{\mw} \right)^{2 \e}}{\eir} \, \ln r -
3 \,  (K_{800} + \e \, K_{801} + \e^2 \, K_{802})
\, \frac{\left( \frac{\m}{\mw} \right)^{2 \e}}{\eir^2} \nonumber \\
&&
-\frac{3}{2} \,  (K_{800} + \e \, K_{801})
\, \frac{\left( \frac{\m}{\mw} \right)^{2 \e}}{\eir} 
\, \left[ 2+ \ln r -4 \ln (\mb/\mw) + 2 i \, \pi  \, \right] 
\nonumber \\
&& + 
h_1 \, \frac{\left( \frac{\m}{\mw} \right)^{4 \e}}{\e}
+ h_2 \, \ln^2 r 
+ h_3 \ln r \ln (\mb/\mw) + h_4 \ln r 
\nonumber \\ 
&& + h_5 \ln
(\mb/\mw) + h_6 \ln^2 (\mb/\mw) + h_7 
\quad .  
\eea 
The first term on the r.h.s. of eq. (\ref{q12tc}) 
is due to infrared singularities coming from the (abelian) graph in set 3
in Fig. 2, where the gluon is radiated from the internal quark; 
the infrared structures appearing in the second and third term
are due to non-abelian diagrams in Fig. 3.
Eq. (\ref{q12tc}) shows that the infrared singularities 
again just multiply
the dimensionally regularized version of the lowest order matrix
element (see eqs. (\ref{lowestfull}) and (\ref{ceps})).

The functions $K_{800}$ and $K_{801}$ appearing in eq. (\ref{q12tc})
are given in eqs. (\ref{k800}) and (\ref{ceps2}). We note that
the function $K_{802}$  is not needed explicitly 
in order to extract $C_{81}$, as we will see later.
The functions $h_i$ in eq. (\ref{q12tc})  read ($z=(m_t/m_W)^2$)
\bea
\label{ht1}
h_1 &=& \frac{z\,(774z^2+810z+144) \, \ln z +137z^5-823z^4+257z^3-425z^2
+958z-104}{72 \, (z-1)^5} - \, \frac{23}{27}  
\nonumber \\
\eea

\be
\label{ht2}
h_2 = \frac{2}{3} \, K_{800} \ , \ 
h_3 = \frac{8}{3} C\, K_{800} \ , \ 
h_4 = -\frac{8}{3} \, K_{800} \ , \ 
h_6 = -6 \, K_{800}  
\ee

\be
\label{ht5}
h_5 =  -
\frac{z\,(162z-72) \, \ln z + 11z^4-110z^3+57z^2+82z-40}{18(z-1)^4} 
+ 6 \, i \, \pi \, K_{800} - 2
\ee
\begin{eqnarray}
h_7 &=&  - \, 
\frac{z(4z^3-40z^2-41z-1) \, \Li}{6(z-1)^4}  
- \frac{8}{3} \, i \, \pi \, K_{800} -\frac{59}{108} \, \pi^2 - 
\frac{185}{324} \nonumber \\
&& - \left[(35964z^3+54756z^2+2592z) \, \ln^2 z \right. \nonumber \\
&&+ \left( 7452z^5-42660z^4-92772z^3-73164z^2+48984z-3360
+3186 \, \pi^2 \, (z^3-z^2) \right) \, \ln z \nonumber \\
&& + (844z^5+40012z^4+90580z^3-148588z^2+16688z+464) \nonumber \\
&& \left. 
+(-885z^5+3363z^4-9381z^3+13629z^2-8142z+1416) 
\, \pi^2
\right] \, /(2592(z-1)^5) \nonumber \\
\end{eqnarray}
\subsection{Counterterms}
As the discussion concerning the counterterms induced by the shifts
in the $t$- and $b$- quark masses is exactly the same as in the
$b \to s \g$ process
in section 3.2, we give immediately the result.
Writing $\delta M_f = W \delta Q_b$ ($f=t,b$) with W given in eq.
(\ref{w}), we get   

\bea
\label{deltaqb}
\delta Q_b &=&  \left\{
-\left( 2\, \ln z + z^2 - 4z + 3 \right) \, \left(
\frac{2}{\e} + 4 \, \ln(\m/\mb) +8/3 \, \right) \right. \nonumber \\
&& \hspace{-1.0cm} \left. +
2  \, \ln^2 z + 2z\,(z-4) \, \ln z
-z^2+8z-7 \, \right\} \, 
\left( \frac{\m}{\mw} \, \right)^{2\e} \,  \frac{ z}{2(z-1)^3} 
\eea

\begin{eqnarray}
\label{deltaqt}
\delta Q_t &=&  
\left\{
\frac{6}{\e} \, \left(
-6z\,(z+1) \, \ln z +z^3+9z^2-9z-1 \, \right) 
+ 18z \, (z+1) \, \ln^2 z  \right. \nonumber \\ 
&& \hspace{-1.0cm} \left. +
(-6z^3-84z^2-18z+24)  \, \ln z +
5z^4-10z^3+126z^2-158z+37 \, \right\} \, \times \nonumber \\
&& \left( \frac{\m}{\mw} \, \right)^{2\e} \,
\frac{z}{3 \, (z-1)^5} 
\end{eqnarray}
Also the counterterms due to the $\sqrt{Z_2}$ factors of the external
quark fields are obtained in the same manner as in section 3.2, 
leading to ($\delta M_{z_2} = W \, \delta Q_{Z_2}$)
\bea
\label{deltaqz}
\delta Q_{Z_2} &=& - 
\left(
\frac{\m}{\mw} \right)^{2\e} \, \frac{4}{3} \, 
\left\{
\frac{2}{\eir} \, \left( K_{800} + \e \, K_{801} \right) \, 
+
\frac{1}{\e} \, \left( K_{800} + \e \, K_{801} \right) \, 
\right. \nonumber \\
&& \left. + \left(4- 6 \, \ln (\mb/\mw) - \frac{3}{2} \, \ln r \, \right) \,
K_{800} \, \right\} \, 
 \quad .
\eea

For the $b \to s g$ case there are additional counterterm
contributions due to the strong coupling constant renomalization
and due to the $\sqrt{Z_3}$ factor associated with the external gluon.
Denoting the combined effect by $\delta M_{g}=W\,\delta Q_g$,
one obtains
\be
\label{deltaqg}
\delta Q_g = \left( -\frac{3}{\e} + f \, \right) \, \left(
K_{800} + \e \, K_{801} \, \right) \quad .   
\ee
As the finite term $f$ will appear also in the corresponding
counterterm in the effective theory, it will drop out when extracting
$C_{81}$.

\section{$b \to s g$ in the effective theory}
\setcounter{equation}{0}
\subsection{Regularized Feynman diagrams}
In the effective theory the matrix element $\hat{M}$ for $b \to s g$ 
is of the form
\be
\hat{M} = \sum_{i=1}^8 \hat{M}^i \quad ; \quad 
\hat{M}^i = \frac{4 i G_F \l_t}{\sqrt{2}} \, C_i \,
\bra s g|O_i|b \ket \quad . 
\ee
We write $\hat{M}^i=
\hat{M}_0^i + \frac{\a_s}{4\p} \, \hat{M}_1^i$
and put the term proportional to $\a_s$
into the form 
\be
\label{effresg}
\frac{\a_s}{4\p} \,
\hat{M}_1^i = W \, \hat{Q}_i \quad , 
\ee
where $W$ is given in eq. (\ref{w}).
As the discussion how to get the quantities $\hat{Q}_i$ is basically identical
as in the $b\to s \g$ case in section 4.1, we just give the results.
Among the four-Fermi operators, only $O_2$ and $O_5$ yield non-vanishing
matrix elements for $b \to s g$.
We get
\be
\label{qhat25}
\hat{Q}_2  \quad , \quad
\hat{Q}_5 = -\frac{2}{9} \, \ln \frac{\m}{\mw} - \frac{1}{6} \,
\tilde{E} \quad ,
\ee
where $\tilde{E}$ is given in eq. (\ref{etilde}). 
Again, we do not have to know $\hat{Q}_2$ explicitly, because
this term also appears in the full theory result; it drops out when
extracting $C_{81}$.

While there is no contribution from the dipole operator
$O_7$, there are various diagrams associated with the operator
$O_8$ (see Figs. 5,6).
The sum of all these contribution is given by
\bea
\label{qhat8}
\hat{Q}_8
&=&  \frac{1}{6} \, C_{80}
\, \frac{\left( \frac{\m}{\mw} \right)^{2 \e}}{\eir} \, \ln r -
3 \,  C_{80} 
\, \frac{\left( \frac{\m}{\mw} \right)^{2 \e}}{\eir^2}
-\frac{3}{2} \,  C_{80}
\, \frac{\left( \frac{\m}{\mw} \right)^{2 \e}}{\eir} 
\, \left[ 2+ \ln r -4 \ln (\mb/\mw) + 2 i \, \pi  \, \right] 
\nonumber \\
&& + C_{80} \, \left(
\frac{11}{3}  \, \frac{\left( \frac{\m}{\mw} \right)^{2 \e}}{\e}
+ 6 \, i \, \pi \, \ln (\mb/\mw) - \frac{8}{3} \, i \, \pi + \frac{2}{3} \,
\ln^2 r - 6 \, \ln^2 (\mb/\mw) 
\right. \nonumber \\
&& \left. 
- \frac{8}{3} \, \ln r + \frac{8}{3} 
\, \ln r \, \ln (\mb/\mw) - \frac{4}{3} \, \ln (\mb/\mw)
+\frac{1}{3} + \frac{59}{36} \, \pi^2
 \right) \, + C_{81} 
\quad .  
\eea 
When comparing with the full-theory expression $Q^{1+2}$ in eq. 
(\ref{q12tc}), one immediately realizes the correspondence of the infrared
singularities. 
To this end it is important that one carefully disentangles
everywhere 
infrared and ultraviolet
poles. Especially, one should  use
the formula
\be
\int \frac{d^dr}{(2\pi)^d} \, \frac{1}{(r^2)^2} = 
\frac{i}{16\pi^2} \, \left(
\frac{1}{\e} - \frac{1}{\eir} \, \right) 
\quad
\mbox{instead of} \quad
\int \frac{d^dr}{(2\pi)^d} \, \frac{1}{(r^2)^2} = 0 \quad .
\ee
An example, where such a situation occurs, is the diagram in Fig 6c.
\subsection{Counterterms}
As the operators mix under renormalization  we have to consider
counterterm contributions induced by operators of the form
$C_i \, \delta Z_{ij} \, O_j$.
We denote their contributions to $b \to s g$ by
\be
\delta \hat{M}_{ij} 
= \frac{4 \, i \, G_F \, \l_t}{\sqrt{2}} \, \bra s g |C_i \,
\delta Z_{ij} \, O_j|b
\ket \quad .
\ee 
The non-vanishing matrix elements
read 
(using $\delta \hat{M}_{ij} = W \, \delta \hat{Q}_{ij}$)
\be
\label{m25glue}
\delta \hat{Q}_{25} = 
- \frac{1}{9} \, \frac{1}{\e} 
\, \left( \frac{\m}{\mb} \, \right)^{2\e} 
\ , \
\delta \hat{Q}_{28} = \frac{19}{27}  \, \frac{1}{\e} \ , \
\delta \hat{Q}_{88} = \, \frac{14}{3} \, \frac{1}{\e} \, C_{80} \quad , 
\ee 
where we made use of the renormalization constants \cite{counterterm}
\be
\label{renconstglue}
\delta Z_{25} = - \frac{1}{9 \e} \, \frac{\a_s}{4\pi}  \quad ,
\quad 
\delta Z_{28} = \frac{19}{27 \e} \, \frac{\a_s}{4\pi}  \quad ,
\quad
\delta Z_{88} = \frac{14}{3\e} \, \frac{\a_s}{4\pi} \, \quad .
\ee 
We note that there are no contributions to $\hat{M}(b \to s g)$ 
from counterterms proportional to evanescent operators.

In analogy to the $b \to s \gamma$ case in section 
4.2, there are the counterterms
from renormalizing  
the $b$-quark mass which explicitly appear in the
definition of the operator $O_8$ 
and from the $\sqrt{Z_2}$ factors for the external quarks.
The counterterm due to the $b$-quark mass renormalization
$\delta \hat{M}_b = W \, \delta \hat{Q}_b$
yields 
\be
\label{deltaqbhat}
\delta \hat{Q}_b = - \frac{4}{3} \, 
\left[ \, \frac{3}{\e} + 6 \, \ln (\m/\mb) \, +4 \,
\right] \, C_{80} \quad ,
\ee
when using the on-shell definition for the $b$-quark mass
(\ref{masscount}), while
the counterterm $\delta \hat{M}_{Z_2}= W \, \delta \hat{Q}_{Z_2}$ 
is given by
\be
\label{deltazgluon}
\delta \hat{Q}_{Z_2} = - \left( \frac{\m}{\mw} \right)^{2\e} \, \frac{4}{3} \, 
\left\{\frac{2}{\eir} \, C_{80} \, 
+
\frac{1}{\e} \, C_{80}  \, 
+ \left(4 - 6 \, \ln (\mb/\mw) - \frac{3}{2} \, \ln r \, \right) \,
C_{80} \, \right\}  \quad .
\ee

Finally , there are counterterms due to the strong coupling constant
renormalization and due to the $\sqrt{Z_3}$ of the external gluon.
As in the full theory, we only give the combined counterterm
$\delta \hat{M}_{g} = W \, \hat{Q}_g$  
\be
\label{deltahatqg}
\delta \hat{Q}_g = \left( -\frac{3}{\e} + f \, \right) \, 
C_{80}   \quad .   
\ee
As  $f$ is the same finite quantity as in the corresponding result 
(\ref{deltaqg}) obtained
in the full theory, we do not need its explicit form,
 because it drops out when extracting $C_{81}$.

\section{Extraction of $C_{81}(\mwt)$}
\setcounter{equation}{0}
To summarize section 6, the order $\a_s$ part $M_1^{ren}$
of the renormalized matrix element 
for $b \to s g$ in the full theory is given by
\be
\label{resrenfullg}
\frac{\a_s}{4\p} \, M_1^{ren} = W \, \left[
Q^{1+2} + \hat{Q}_2 + \delta Q_b + \delta Q_t + \delta Q_{Z_2}
+ \delta Q_{g} \,
\right] \quad ,
\ee
where the quantities in the bracket on the r.h.s. of 
eq. (\ref{resrenfullg})
are given in eqs. (\ref{q12tc}), (\ref{identg}), (\ref{deltaqb}),
(\ref{deltaqt}), (\ref{deltaqz}) and (\ref{deltaqg}),respectively; 
the prefactor $W$ is given in eq. (\ref{w}).

The corresponding renormalized matrix element in the effective theory
can be obtained from the information in section 7; $\hat{M}_1^{ren}$
reads
\be
\label{resreneffg}
\frac{\a_s}{4\p} \, \hat{M}_1^{ren} = 
W \, \left[\hat{Q}_2 + \hat{Q}_5 +
 \hat{Q}_8 +
\delta \hat{Q}_{25} + \delta \hat{Q}_{28} 
+ \delta \hat{Q}_{88} 
+ \delta \hat{Q}_b  + \delta \hat{Q}_{Z_2} +\delta \hat{Q}_g \,
\right] \quad ,
\ee
where the various quantities in the bracket are given in eqs.
(\ref{qhat25}), (\ref{qhat8}),
(\ref{m25glue}), (\ref{deltaqbhat}), (\ref{deltazgluon})
and (\ref{deltahatqg}). 

Before we extract $C_{81}$, which enters $\hat{M}_1^{ren}$
via $\hat{Q}_8$ (see eq. (\ref{qhat8})), we
should point out 
that the discussion concerning the infrared singularities
is similar as in the $b \to s \g$ case in section 5;
all the formulae are written in such a way
that we simply can discard 
the terms proportional to the poles in
$\eir$ in both versions of the theory.
The extraction of $C_{81}$ is then straightforward. 

To summarize: Writing the
Wilson coefficient $C_8(\mwt)$ at the matching scale
$\mwt$ in the form
\be
\label{resc8}
C_8(\mwt) = C_{80}(\mwt) + \frac{\a_s}{4\p} \, C_{81}(\mwt) \quad ,
\ee
we obtain (in the naive dimensional regularization scheme)
\bea
\label{resc81}
C_{81}(\mwt) &=& -\frac{z\,(4z^3-40z^2-41z-1)}{6(z-1)^4} \, \Li
- \frac{z^2\,(17z+31)}{2(z-1)^5} \, \ln^2 z \nonumber \\
&& - \frac{210z^5-1086z^4-4839z^3-3007z^2+2114z-304}{216(z-1)^5}
\, \ln z \nonumber \\
&& + \frac{611z^4-13346z^3-29595z^2+1510z-652}{1296(z-1)^4} +
\nonumber \\
&& \hspace{-0.2cm}
+ \frac{z^2\,(17z+31)}{(z-1)^5} \, \ln z \, \ln \frac{\mwt}{\mw} +
\frac{89z^4-446z^3-1437z^2-950z+152}{54(z-1)^4} \, \ln \frac{\mwt}{\mw} 
\quad . 
\nonumber \\
\eea
Here, $z=(m_t(\mwt)/m_W)^2$, where $m_t(\mwt)$ is the 
$\overline{MS}$ top quark mass at the renormalization scale $\mwt$.
The lowest order function $C_{80}$ is given in eq. (\ref{k800}).

Taking into account that the result of Adel and Yao \cite{Adel}
is given in the so-called $R^*$ renormalization scheme,
our result is identical. 

\section{Summary}
\setcounter{equation}{0}
The order $\a_s$ corrections
to the Wilson coefficients $C_7$ and $C_8$
are a very crucial ingredient for the prediction of the branching
ratio for $b \to X_s \gamma$ in  next-to-leading logarithmic precision.
As these corrections, which involve 
many two-loop diagrams in the full theory, have been calculated so far
by one group \cite{Adel} only, we presented in this work a detailed
recalculation. We extracted the $O(\a_s)$ corrections to $C_7$ and 
$C_8$ by comparing the on-shell processes $b \to s \g$ and $b \to s g$
in both versions of the theory. We evaluated the two-loop integrals 
in the full theory by using the Heavy Mass expansion method. 
Our $\a_s$ corrections ($C_{71}$ and $C_{81}$) to the Wilson coefficients 
$C_{7}$ and
$C_{8}$ completely agree with the findings of Adel and Yao.

We should point out that our result (as well as Adel and Yao's) for
$C_{71}(\mwt)$ and $C_{81}(\mwt)$ is \`a priori specific to the basis
given in eq. (\ref{operators}). However, the same answer is obtained for 
these Wilson coefficients
when working in the basis recently used by Chetyrkin, Misiak and
M\"unz. \\

\vspace{2cm}

{\bf Acknowlegdments}
\newline
We thank A. Ali, M. Misiak, U. Nierste and S.J. Rey for helpful 
discussions.

\newpage

\section*{Figure Captions}
\subsection*{Figure 1}
\begin{description}
\item[a] Lowest order diagram for $b \to s \g$ in the full-theory.
A cross denotes a possible location where the photon can be emitted.
The wavy line stands for a $W$ or unphysical Higgs boson ($\Phi$).
In the $b \to s g$ case the cross at the $W/\Phi$ has to be ignored. 
\item[b] Typical two-loop graph for $b \to s \g$.
\item[c-e] Subdiagrams of b) which contribute in the Heavy Mass Expansion.
See text.
\end{description}
\subsection*{Figure 2}
Complete list of two-loop diagrams for $b \to s \g$ in the full-theory. 
A cross corresponds
to a possible location for the photon emission. \\
For the $b \to s g$ process, this figure is a complete list of diagrams
not involving the gluon triple coupling. (In the $b \to s g$ case
the crosses at the wavy ($W/\Phi$) lines should be ignored.)
\subsection*{Figure 3}
Complete list of two-loop diagrams involving the triple gluon vertex
(for the $b \to s g$ process).
\subsection*{Figure 4}
Diagrams associated with the operators $O_7$ and $O_8$ in the effective
theory for $b \to s \g$. See text.
\subsection*{Figure 5}
Abelian diagrams associated with the operator $O_8$ in the effective
theory for $b \to s g$. See text.
\subsection*{Figure 6}
Non-Abelian diagrams associated with the operator $O_8$ in the effective
theory for $b \to s g$. See text.

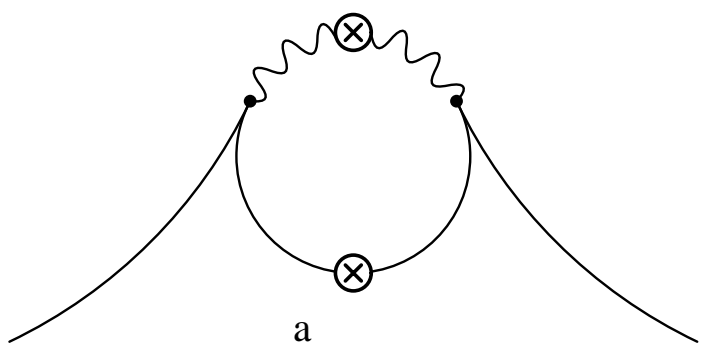
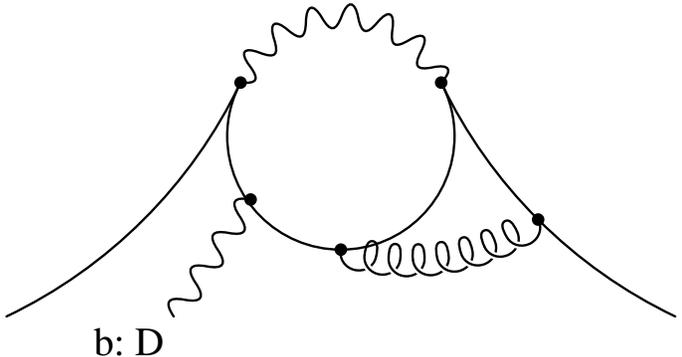
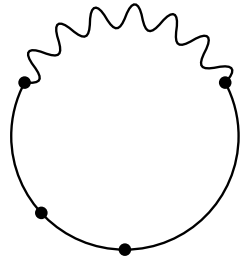
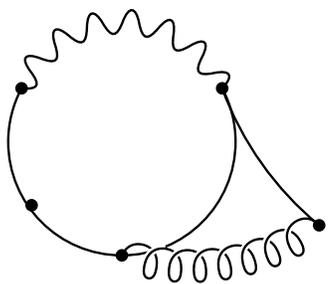
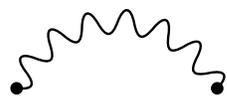
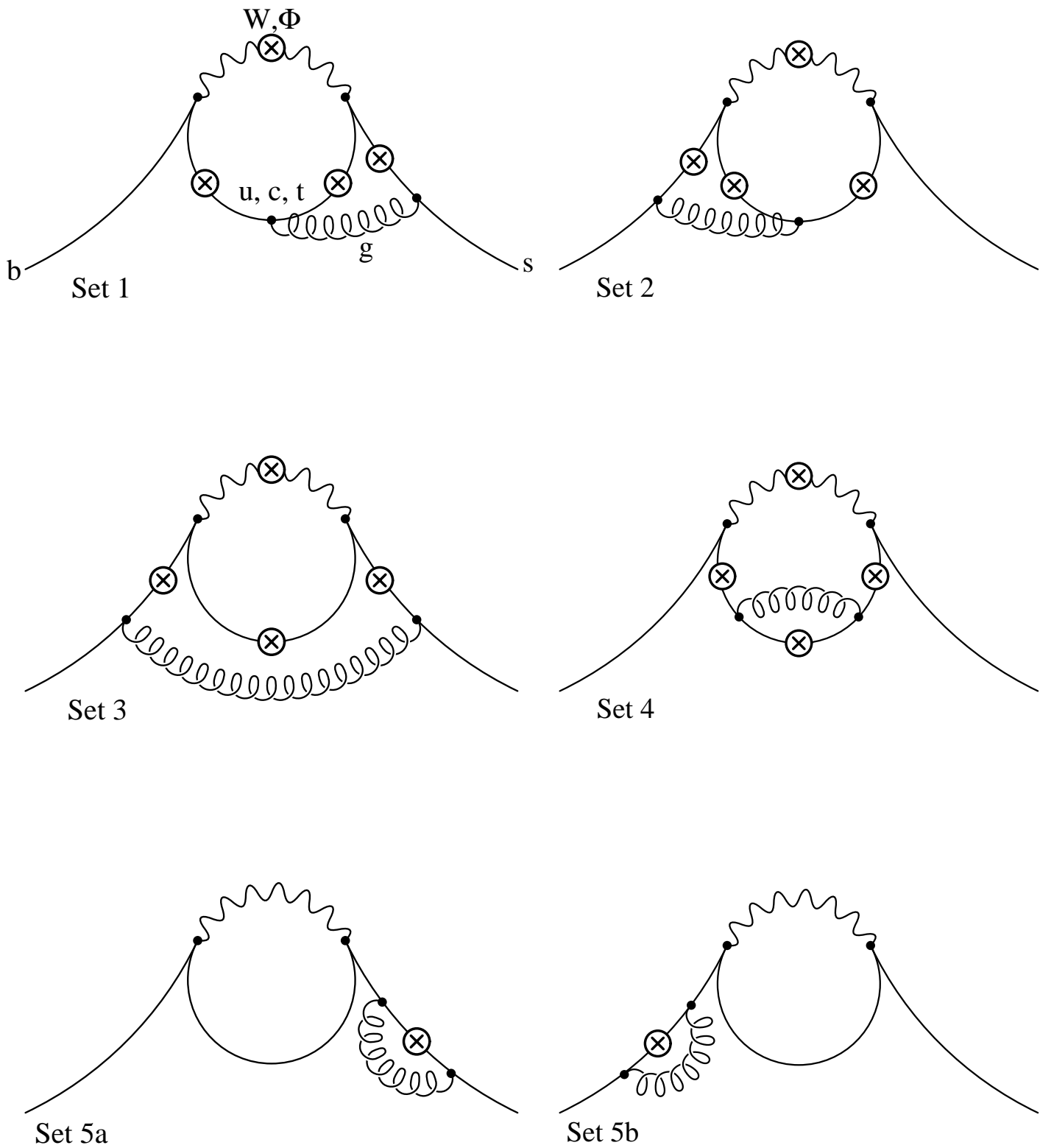
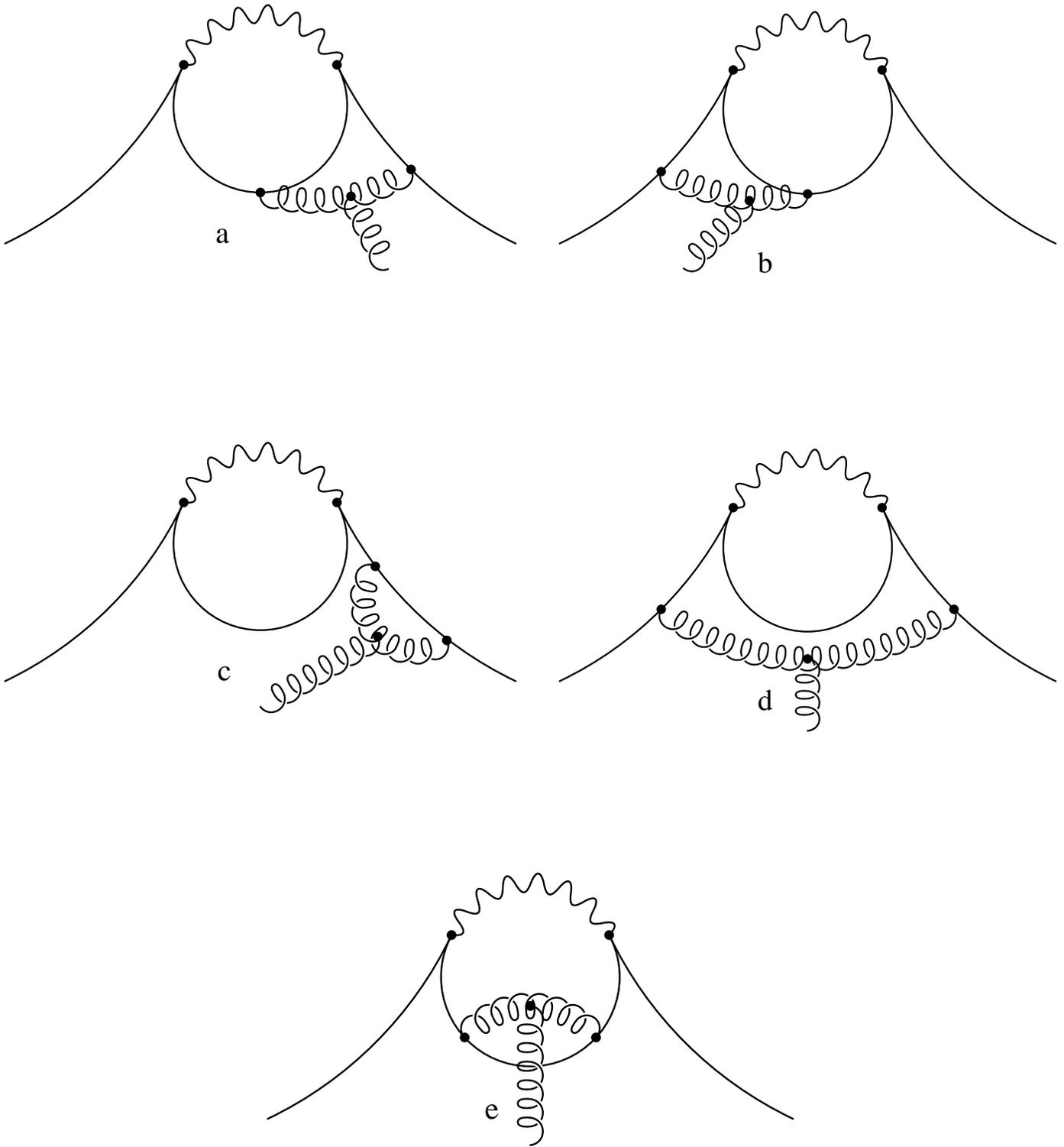
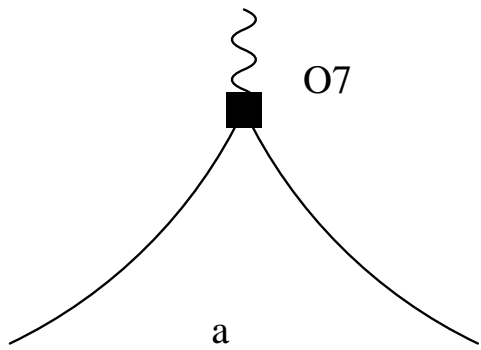
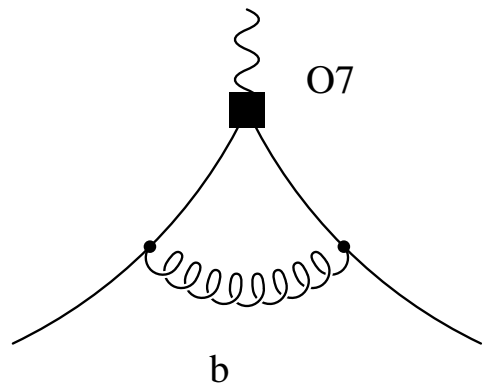
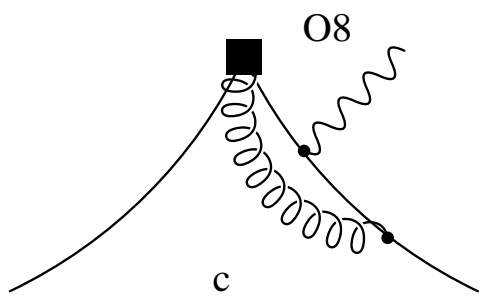
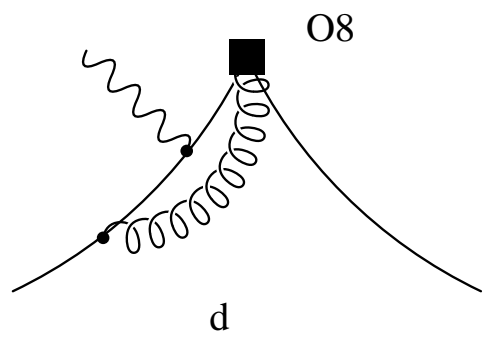
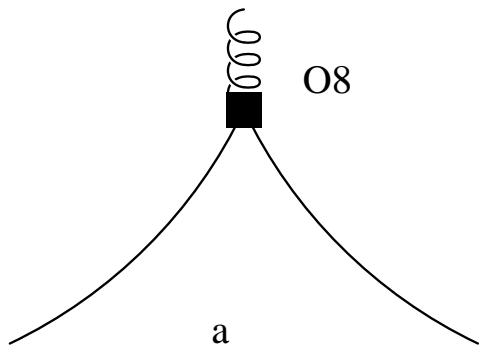
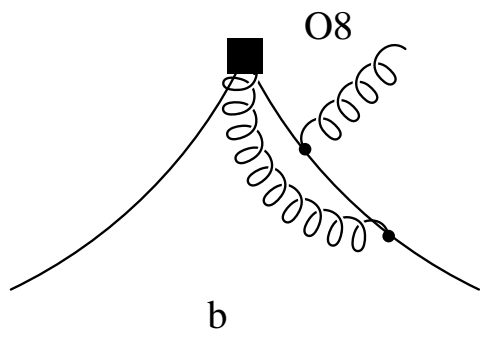
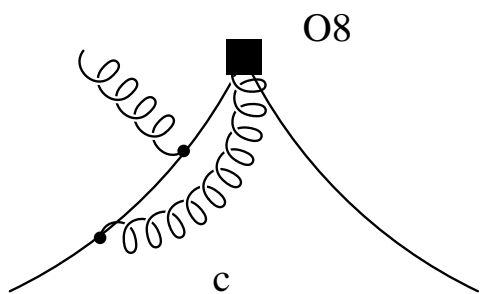
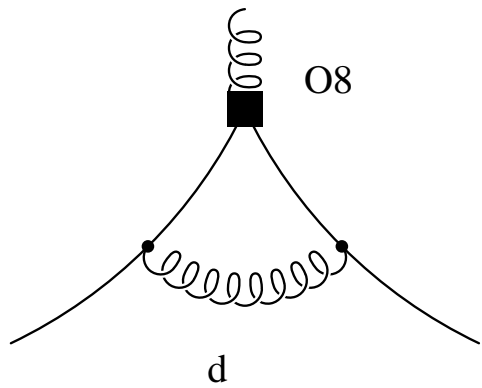
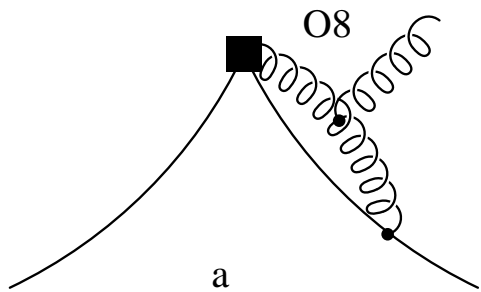
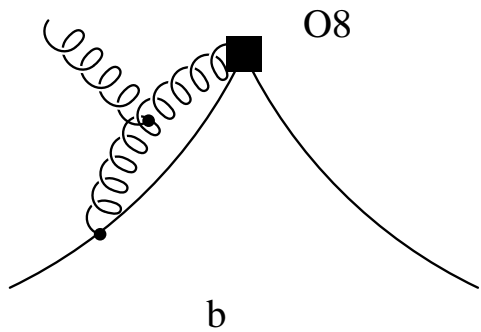
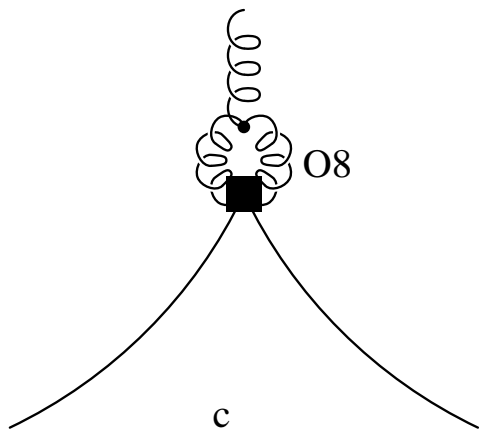
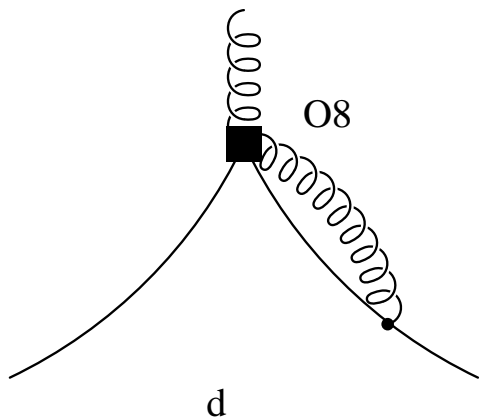
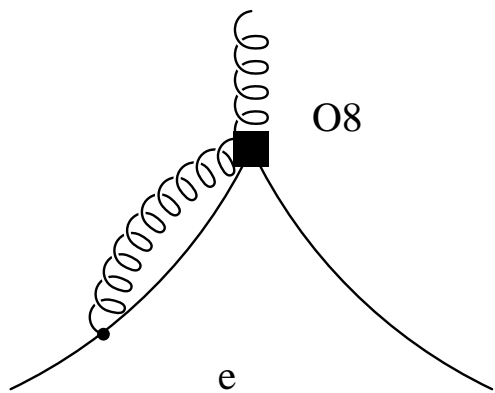
\end{document}